\documentclass{article}
\usepackage{graphicx} 
\usepackage{float}
\usepackage{subcaption}
\usepackage{graphicx}
\usepackage{cuted}
\usepackage{url}
\usepackage{booktabs}
\usepackage{multirow}
\usepackage{amsmath}  
\usepackage{amssymb}  
\usepackage[colorlinks=true, allcolors=blue]{hyperref}

\usepackage[margin = 2.5cm]{geometry} 

\title{Testing a cosmological teleparallel modified \textit{f(T)} model with parametric \textit{H(z)}}

\author{A. O. Aquino and J.E.G. Silva} 
\date{september 2026}

\begin{document}

\maketitle
\begin{abstract}
We study the late-time cosmological expansion of a modified teleparallel gravity model. This modified gravitational lagrangian yields a cosmological constant term along with power-law corrections to the Teleparallel Equivalent of General Relativity (TEGR) for small $\lambda$. By combining observational data from cosmic chronometers, Type Ia supernovae from the Pantheon+ dataset, and Baryon Acoustic Oscillations (BAO) from DESI DR2, we constrain the parameters of the modified gravitational dynamics assuming a quadratically parameterized Hubble rate $H(z)$. The joint observational fit indicates a late-time accelerated expansion with a deceleration parameter of $q_0 = -0.388_{-0.038}^{+0.038}$. Furthermore, we analyze the effective energy density, pressure, and equation-of-state parameter $\omega$, showing that this modified gravity theory naturally accommodates quintessence-like and phantom-like behavior. Finally, a linear scalar perturbation analysis confirms that the cosmological background is dynamically stable against homogeneous fluctuations.
\end{abstract}

\section{Introduction}
The $\Lambda CDM$ model is maintained to this day with the status of ``standard cosmological model'' due to its success in relation to observational data \cite{BULL201656}. This model takes into account the concepts of cold dark matter (CDM) to explain the rotation speed of galaxies \cite{Zatrimaylov:2021ijd}, and dark energy to explain the accelerated expansion of the universe. The latter concept is characterized in the model by the cosmological constant $\Lambda$ \cite{Perivolaropoulos:2021jda}. However, despite its great success with observations, the cosmological constant brings with it some problems. Indeed, since its source, known as the dark energy, has an equation of state  $p=-\rho$, it leads us to question whether there is a better explanation of who is responsible for the cosmic expansion. 


The $\Lambda CDM$ model has two main pillars, gravity from the General Theory of Relativity (GRT) \cite{Einstein:1917ce} and the Cosmological Principle (CP) \cite{Perivolaropoulos:2021jda}, which states that the universe is homogeneous and isotropic for large scales. These bases, when solving the GRT field equations for the Friedmann-Lemaitre-Robertson-Walker metric, we obtain the Friedmann equations, equations that dictate the cosmic dynamics in expansion based on a curved space where $R\neq 0$ and a scale factor a(t) \cite{Friedman:1922kd,Friedmann:1924bb,Lemaitre:1927zz,Robertson01051928,Walker:1937}. However, there is an alternative approach, we can modify the geometry of the problem, and instead of working with a metric field $g_{\mu\nu}$, we work with a field of tetrads referring to a flat tangent space where $R=0$ and non-zero torsion ($T \neq 0$). This approach is called Teleparallelism Equivalent to General Relativity (TERG) \cite{Cai:2015emx}. It is important to make it clear that although they can reproduce the same results under certain conditions, TERG is not equivalent to TRG, but rather an alternative theory.

There are currently several possibilities to specifically circumvent the problem of cosmic expansion, for example, modifications in the matter and energy content of the universe, such as the $\omega$CDM models \cite{michele:01}, scalar field quintessence models \cite{Leon:2021lct,ROY2022101037} and models by adopting a scalar field with phantom behavior  \cite{ROY2022101037,Avsajanishvili:2023jcl}. Such models propose that the responsible for the cosmic expansion is an exotic fluid or scalar field which will have a dynamical state parameter $\omega$ dependent on the redshift z. Another way to approach a possible alternative explanation for the problem of dark energy is to propose modifications in the gravitational theory of the cosmological model, as do the theories f(R) \cite{Clifton:2011jh}, f(Q) \cite{MYRZAKULOV2023101268} and f(T) \cite{doi:10.1142/9789813226609_0074}. Such modifications are performed in the Einstein-Hilbert action so that instead of depending only on the curvature scalar R or torsion scalar T, it will depend on a function of them. The last proposal cited is based on models in which cosmic expansion is explained by gravitational theory itself, without the need to add an exotic fluid. In this way, we will have an effective behavior and not a fluid per se, which can alleviate inconsistencies when we find negative pressure behavior. One of the major problems associated with f(R) models is that the solutions to the field equations tend to be high-order differential equations \cite{SINGH2024101658}, which make it difficult to find analytical solutions, and it is often necessary to find the solution using numerical techniques. An advantage of $f(T)$ models over $f(R)$ models is the fact that $f(T)$ gravity yields equation of motion up to the second order in derivatives \cite{Capozziello:2018hly}, which avoids some unstable solutions.

Some $f(T)$ models have already been tested, such as exponential and power law \cite{PhysRevD.82.109902} Lagrangian, which, given the observational data, we can state that they have great potential to provide plausible explanations for the results obtained. In this work, we will first review gravitation from the point of view of teleparallelism, how to make it equivalent to TRG and how to modify it. Then, we will discuss the cosmological solutions obtained from the solution of the teleparallelism field equations for a Friedmann-Lemaître-Robertson-Walker metric, specify the modification to be made in the gravitational sector, test it against observational data and discuss the results obtained.

The work is organized as the following. In the section \ref{sec2}, the basics of the teleparallel $f(T)$ models are reviewed. In the section \ref{sec3}, the modified Friedmann equations and the respective effective energy density and pressure are obtained in a general teleparallel $f(T)$ gravity. In section \ref{sec4}, the logarithmic model is introduced and an $H(z)$ parametrization is adopted for a model-independent qualitative analysis, which does not constitute a direct solution of the $f(T)$ field equations. In section \ref{sec5}, the $H(z)$ parametrization is tested adopting three distinct datasets (Hubble, Supernova type IA and BAO). In the section \ref{sec6}, we use the data from the $H(z)$ obtained in the section \ref{sec5} to obtain the effective energy density, pressure and the equation of state. In the section \ref{sec7}, the dynamical stability analysis of the cosmological background under linear scalar perturbations is performed.

\section{Teleparallel modified gravity models}
\label{sec2}

Teleparallel gravitational theories are  alternative theories of gravitation based on the spacetime torsion, instead of working with a curvature scalar. For this, we will use the tetrads $e_{a}(x^{\mu})$, where $a=0,1,2,3$ describe vectors that form an orthonormal basis in a tangent space for each point $x^{\mu}$ of a manifold. The vectors $e_{a}$ are orthogonal to each other, so that $e_{a} \cdot e_{b} = \eta_{ab}$, where $\eta_{ab}$ is the Minkowski metric of the local tangent space given by $\eta_{ab} = diag(-1,1,1,1)$, that is, for each point $x^{\mu}$ in a curved space, we will have a correspondence in a flat tangent space.

Each tetrad $e_{a}$ has components $e_{a}^{\mu}$, where $\mu = 0,1,2,3$ represent coordinates of the curved spacetime, thus we can express the vector as $e_{a}= e^{\mu}_{a}\partial_{\mu}$, thus the Latin indices will be components of the tangent spacetime and the Greek indices will be components of the curved manifold. As in the usual approach of the Theory of General Relativity, we need a metric, we can express it as $g_{\mu\nu}= \eta_{ab} e^{a}_{\mu}e^{b}_{\nu}$, thus we have a direct relationship between the metric of the curved manifold and the flat metric of the tangent spacetime. We will also have that the inverse will obey the relation $e^{a}_{\mu}e^{\mu}_{b} = \delta^{a}_{b}$, where $\delta^{a}_{b}$ is the Kronecker delta. By the metric too, we can define that $\sqrt{-g}= det[e^{a}_{\mu}] = e$. If we are changing the way of describing gravitational effects in space-time, it is clear that there will also be changes in the way of writing the connection $\Gamma$, which in the most usual way, it is common to use the symmetric part of the general connection, which we know as the Levi-Civita connection, however, based on the tetrad postulate $\nabla_\nu e^a_\mu = 0$, our connection is given by Eq.(\ref{weitzenbock}) \cite{Bahamonde:2021gfp} 
\begin{equation}
\label{weitzenbock}
    \Gamma^\lambda_{\mu\nu} = e^\lambda_a \left( \partial_\nu e^a_\mu + \omega^a_{\ b\nu} e^b_\mu \right).
\end{equation}
Accordingly, the torsion tensor has the form \cite{Krssak:2015oua}
\begin{equation}
\label{torsion}
    T^\lambda_{\ \mu\nu} = \Gamma^\lambda_{\ \mu\nu} - \Gamma^\lambda_{\ \nu\mu} = e^\lambda_a \left( \partial_\mu e^a_\nu - \partial_\nu e^a_\mu + \omega^a_{\ b\mu} e^b_\nu - \omega^a_{\ b\nu} e^b_\mu \right).
\end{equation}
From Eq.(\ref{torsion}) we can define the so-called torsion scalar, as
\begin{equation}
    T = T_{\mu\nu}^{\lambda}S_{\lambda}^{\mu\nu},
\end{equation}
where the tensor $S^{\mu\nu}_{\lambda}$ is defined as
\begin{equation}
   S^{\mu\nu}_{\lambda} = \frac{1}{2}\left (K^{\mu\nu}_{\lambda} + \delta^{\mu}_{\lambda}T^{\sigma \nu}_{\sigma} - \delta^{\nu}_{\lambda} T^{\sigma \mu}_{\sigma}\right )
\end{equation}
And the tensor $K^{\mu\nu}_{\sigma}$ is called the contortion tensor, defined as the difference between the Weitzenböck and Levi-Civita connections, respectively, as represented in Eq.~\eqref{contorsion}
\begin{equation}
\label{contorsion}
    K^{\mu\nu}_{\lambda} = \Gamma^{\lambda}_{\mu\nu} - ^{\circ}\Gamma^{\lambda}_{\mu\nu}
\end{equation}
It is worth noting that it is also possible to obtain the torsion scalar from contractions of indices in products with the torsion tensor itself, which can be an alternative to the path of calculating the contortion $K$ and the tensor $S$ \cite{Bamba:2013jqa}.

Before proceeding, it is important to mention that this is the covariant formulation of teleparallelism \cite{Krssak:2015oua, Bejarano:2019fii, Golovnev:2017dox, Tamanini:2012hg}, in which the spin connection $\omega^a{}_{b\nu}$ is incorporated into the teleparallel connection. In this way, the tetrad and the spin connection transform jointly under local Lorentz transformations, ensuring the covariance of the formulation and eliminating the frame dependence associated with the pure-tetrad formulation. In the latter, the so-called Weitzenböck gauge is adopted, in which the spin connection is chosen to vanish, $\omega^a{}_{b\nu}=0$, causing the teleparallel connection to reduce to the Weitzenböck connection, given by $\Gamma^\lambda_{\mu \nu} = e^\lambda_a \partial_\nu e^a_\mu$.

To obtain the field equations, just as we do in the curvature approach of General Relativity, we will use the Einstein-Hilbert action, and here we can start making modifications to the gravitational sector to obtain alternative results, so we can write the action as Eq.~\eqref{action_mod}
\begin{equation}
\label{action_mod}
    S = \frac{1}{2 \kappa^2}\int ed^{4}x (T + f(T)) + \int ed^{4}x (\mathcal{L}_{m}+\mathcal{L}_{r}),
\end{equation}
where $\mathcal{L}_{m}$ and $\mathcal{L}_{r}$ are the matter and radiation Lagrangians, respectively. The field equation will be given by Eq.~\eqref{field_mod} \cite{Krssak:2015oua} 
\begin{equation}
\label{field_mod}
\begin{aligned}
    e^{-1} f_T \partial_\nu \left(e S_a^{\ \mu\nu}\right) &+ f_{TT} S_a^{\ \mu\nu} \partial_\nu T - f_T T^b_{\ \nu a} S_b^{\ \nu\mu} 
+ f_T \omega^b_{\ a\nu} S_b^{\ \nu\mu} + \frac{1}{2} f(T) e_a^\mu 
= \kappa^2 e_a^\rho \mathcal{T}_\rho^\mu,
\end{aligned}
\end{equation}
where $f_{T}(T) = \partial f(T)/\partial T$, $f_{TT}(T) = \partial^{2}f(T)/\partial T^{2}$ and $\mathcal{T}^{\nu}_{\mu}$ is the energy-momentum tensor. Furthermore, we have that $\kappa^2 = \frac{8 \pi G }{c^4}$. It is important to emphasize that, when considering $f(T) = 0$, Eq.\eqref{action_mod} will be reduced to the usual field equation of teleparallelism equivalent to General Relativity. It is also relevant to remember that the solution of Eq.\eqref{field_mod} is directly linked to the choice of tetrads for the model, thus, an appropriate set of tetrads can provide an adequate solution.

\section{$f(T)$ Cosmology}
\label{sec3}

In order to construct a homogeneous and isotropic cosmological model, let us consider the Friedmann-Lemaître-Robertson-Walker (FRLW) metric with a line element
\begin{equation}
\label{frwmetric}
    ds^{2} = - dt^2 + a^{2}(t)\delta_{ij}dx^{i}dx^{j},
\end{equation}
where $a(t)$ represents the dimensionless  scale factor, derived from Hubble's law \cite{hubble:1929}. From the metric above we can express the tetrad in a diagonal form $ e_{\mu}^{i} = \text{diag}(1, a(t), a(t), a(t))$ \cite{Bamba:2013jqa}. The corresponding torsion scalar has the form
\cite{Capozziello:2018hly,Bamba:2013jqa}
\begin{equation}
\label{torsion_scalar}
    T = -6H^{2},
\end{equation}
where $H$ is the Hubble parameter, defined as $ H = \frac{\dot{a}}{a}$.

Adopting a perfect fluid as a source, the energy-momentum tensor is given by
\begin{equation}
T_{\mu\nu} = (\rho + p)u_{\mu}u^{\mu} \eta_{ij} e^{i}_{\mu} e^{j}_{\nu} + p \eta_{ij} e^{i}_{\mu} e^{j}_{\nu},
\end{equation}
where $u_{\mu}$ is the 4-velocity, which satisfies the condition $u^{\mu}u_{\mu} = -1$, and $\rho$ and $p$ represent, respectively, the energy density and pressure of the fluid. Moreover, for an isotropic source, we can consider the equation of state $p = \omega \rho$. 

Using Eq.~\eqref{torsion_scalar} and the metric of Eq.~\eqref{frwmetric} in Eq.~\eqref{field_mod}, and taking into account the energy-momentum tensor described above, we obtain the modified Friedmann equations \cite{Krssak:2015oua, Sandoval-Orozco:2023pit}
\begin{equation}
\label{fried_mod}
    3H^{2} + Tf_{T}(T) - \frac{f(T)}{2} = \kappa^{2} \rho,
\end{equation}
\begin{equation}
\label{fried_mod2}
     \dot{H}\left(1-f_{T}(T) -2Tf_{TT}(T)\right) = -\frac{\kappa^2}{2}(\rho + p).
\end{equation}
Comparing equations \eqref{fried_mod} and \eqref{fried_mod2} with the Friedmann equations obtained from the $\Lambda CDM$ model, which would be $3H^2 = \rho$ and $2\dot{H} + 3H^2 = -p$ (considering the other constants as equal to 1), it is possible to observe that the extra terms (terms that contain $f(T)$ and its derivatives) come from the sum of $f(T)$ in the action, thus, we can interpret it as an effective fluid with effective energy density and pressure. In this way, it is possible to interpret these effective terms as an interpretation for what we call the dark energy responsible for cosmic expansion, but the effect has a purely geometric origin. We then write the effective energy density as
\begin{equation}
\label{rhog}
    \rho_{g} = -\frac{1}{2\kappa^{2}}(2Tf_{T}(T)- f(T)),
\end{equation}
whereas the effective pressure as
\begin{equation}
\label{pg}
    p_{g}= -\frac{1}{2 \kappa^2}(-8T\dot{H}f_{TT}(T) - f_{T}(T)(-2T + 4\dot{H})+f(T))
\end{equation}
Using the relation between $\rho$ and $p$ of the perfect fluid discussed earlier $p = \omega \rho$, we have that the effective state parameter for this perfect fluid will be \cite{Bamba:2010wb,Capozziello:2018hly}
\begin{equation}
\label{omegag}
    \omega_{g} = \frac{-8T\dot{H}f_{TT}(T) - f_{T}(T)(-2T + 4\dot{H})+f(T)}{2Tf_{T}(T)- f(T)}
\end{equation}
In many cosmological models, the effective equation of state, $\omega_{g} = \frac{p_{g}}{\rho_{g}}$, plays a central role in describing the accelerated expansion of the universe. In the context of modified gravity theories, such as those based on the function $f(T)$, where $T$ is the torsion scalar, the cosmological dynamics can be fitted according to the specific forms adopted for $f(T)$ and the Hubble function, $H(t)$. When the effective equation of state satisfies $-1 < \omega_{g} < -\frac{1}{3}$, the universe exhibits quintessential behavior. In this regime, dark energy behaves analogously to a canonical scalar field with sufficient negative pressure to drive the accelerated expansion of the universe, without violating the energy condition. This behavior is widely discussed in the cosmological literature and is in agreement with observations of an accelerating expanding universe. On the other hand, if $\omega_{g} < -1$, the cosmological behavior is of the phantom type. In this case, the equation of state implies the violation of the weak energy condition, resulting in a superaccelerated expansion phase. This scenario is particularly interesting because it can lead to extreme consequences, such as the ``Big Rip'', in which the expansion of the universe grows so rapidly that, in the distant future, even atomic and subatomic structures would disintegrate. Thus, the appropriate choice of the functions $f(T)$ and $H(t)$ is fundamental to determine the cosmological behavior of the universe, because these choices directly affect the effective equation of state and, consequently, the evolution of the universe in different dark energy regimes. Thus, depending on the form adopted for $f(T)$, the model can describe either a quintessence regime or a phantom regime, agreeing with the hypothesis of an accelerating expanding universe observed in recent decades. To better situate this effective equation of state within observational cosmology, it is instructive to relate $w_g$ to the broader $w\mathrm{CDM}$ framework commonly used to search for departures from the standard $\Lambda\mathrm{CDM}$ model ($w = -1$). While observational constraints often probe dark energy using phenomenological parametrizations of a dynamical equation of state $w(z)$ such as the widely used Chevallier-Polarski-Linder (CPL) parametrization, $w(z) = w_0 + w_a \frac{z}{1+z}$ the $f(T)$ framework provides a physical mechanism for such evolution. In this context, the effective parameter $w_g(t)$ given by Eq.~\eqref{omegag} plays the role of an evolving dark energy parameter $w(z)$ whose dynamics are directly governed by the functional form of $f(T)$ and $H$, rather than an ad-hoc ansatz. Consequently, mapping $w_g(z)$ onto $w\mathrm{CDM}$ parameter spaces allows one to directly test whether specific $f(T)$ models naturally accommodate quintessence-like ($w_g > -1$), phantom-like ($w_g < -1$), or phantom-divide crossing behaviors in light of late-time observational data.

\section{A logarithmic $f(T)$ cosmological model}
\label{sec4}

In the previous section, we showed how the particular $f(T)$ function modifies the Friedman equations.
There are many possible choices for the form of $f(T)$, for instance a power law $f(T) = \alpha (-T)^{n}$ \cite{Linder:2010py}, an exponential $f(T) = T_{0} \left ( 1- Exp\left[-\alpha\sqrt{T/T_{0}} \right ] \right )$ \cite{Linder:2010py}, or a logarithm $f(T) = D \log(b T)$ \cite{MANDAL2020100551}. The parameters 
introduced in the $f(T)$ are expected to be such that reproduce the well-known results of general relativity and also provide a gravitational origin of the late expansion.

The construction of this $f(T)$ model was, admittedly, driven primarily by mathematical considerations rather than a direct physical motivation; nonetheless, the underlying inspiration stems from power-law $f(T)$ models, which do carry well-established physical motivations in the literature. Building on previous works that introduce powers of the torsion scalar, such as the power-law considered in ref.~\cite{Linder:2010py}, in which an extra parameter controls how influential this modification becomes in the dynamical equations, we propose here not a single power but an infinite sum of powers of $\lambda T$ with a decreasing weight $1/k$ at each order $k$. The motivation for this choice is purely constructive: a series of this type admits an exact analytical closed form in terms of a logarithm, allowing infinitely many higher-order corrections to be condensed into a single closed expression, rather than truncating the series at some arbitrary order. Within this construction, our $f(T)$ can be written as
\begin{equation}
    f(T) \propto \left(\lambda T + \frac{(\lambda T)^2}{2} + \frac{(\lambda T)^3}{3} + \dots\right)^n,
\end{equation}
where the parameter $n$ plays the same role as in ref.~\cite{Linder:2010py}: it controls the overall weight of the modification. In particular, $n = 0$ recovers $f(T) = \text{const.}$, reducing the model to the standard cosmological scenario. Using the identity $\sum_{k=1}^{\infty} x^k/k = -\ln(1-x)$, the series above condenses exactly into
\begin{equation}
\label{lag1}
    f(T) = 2\Lambda \left(\log_\gamma(1-\lambda T)\right)^n,
\end{equation}
where $\Lambda$, $n$ and $\lambda$ are constants, and the logarithm base $\gamma$ absorbs the normalization of the series. This choice avoids divergences in the $T \rightarrow 0$ limit only insofar as the argument $(1-\lambda T)$ remains positive over the physical domain considered. This closed form leads to interesting behaviors: for $n=1$ and $\lambda T \rightarrow 0$, with $\gamma$ constant, the function $f(T)$ can be written as the power-series
\begin{equation}
    f(T) = 2\Lambda\left(\frac{\lambda T}{\ln\gamma} + \frac{\lambda^2T^2}{2\ln\gamma} + \dots\right).
\end{equation}
Note that both $\Lambda$ and $\lambda$ have mass dimension $[\Lambda]=[\lambda]=M^2$. Considering Eq.~\eqref{lag1}, we obtain a positive energy density $\rho_{g}$ and a negative pressure $p_{g}$, just like a cosmological constant solution, but here $\omega$ will be dynamic. For this work, we consider $\gamma = 2 - \lambda T$, so the expansion in powers of $\lambda T$ will be expressed by Eq.~\eqref{expansion1}
\begin{equation}
\label{expansion1}
    -log_{(2-\lambda T)}\left ( 1-\lambda T \right ) = \frac{\lambda T}{ln (2)} + \frac{\lambda^2 T^{2}(1+ln(2))}{2(ln (2))^2}+...
\end{equation}
This hypothesis is based on an approximately "constant" behavior, if both the argument and the base of the logarithm change gradually and proportionally, the logarithm can be considered effectively constant, since the variations in argument and base occur slowly. This occurs because a logarithm with a dynamic argument and base can always be rewritten as a quotient of natural logarithms. If we want a slow contribution of $f(T)$ to the GR part of our action (i.e., only $T$), we must choose a very small value of $\lambda$. In this work, we choose $\lambda = 10^{-3}$, as this value preserves the function from exhibiting chaotic behavior.

From this point on, it becomes necessary to adopt an ansatz in order to analyze the evolution of the cosmic expansion of the universe, as well as to fit the constant parameters $\alpha$ and $\beta$ of the function $f(T)$. In the literature, there are well-established justifications for employing this kind of assumption~\cite{MYRZAKULOV2023101268}. This technique is commonly referred to as a model-independent analysis, since, in general, the Hubble parameter can be modeled in an arbitrary manner, provided that the choice is physically consistent. By considering a parametrization of $H(z)$ in these terms, we can reproduce an universe in accelerated expansion and perform tests on any modified model.

Once a functional form for the Hubble parameter $H$ is assumed, it becomes possible to derive other relevant cosmological parameters, such as the deceleration parameter $q$ and the jerk parameter $j$, which may be useful for specific types of analysis. The deceleration parameter $q$ is self-explanatory by its name, however, its sign is of particular importance: negative values indicate an accelerated expansion of the universe, while positive values correspond to a decelerated expansion. The jerk parameter $j$, in turn, is related to the third time derivative of the scale factor and is defined as $j = \frac{d^{3}a(t)}{dt^3}$~\cite{AlMamon:2018uby}. Nevertheless, this work will focus on the construction based on the Hubble parameter and a brief analysis of the deceleration parameter, so the jerk parameter will not be considered. In this context, we adopt the following parametrized form for the Hubble parameter as a function of the redshift $z$ :
\begin{equation}
\label{parametricH}
    H(z) = H_{0}(1 + \alpha z + \beta z^{2}),
\end{equation}
where $H_{0}$ is the Hubble constant, and $\alpha$ and $\beta$ are the same constants appearing in the function $f(T)$ given by Eq.\eqref{lag1}. These will act as free parameters in a forthcoming numerical analysis. This parametrization is consistent in the limit $z \rightarrow 0$, since we recover $H(0) = H_{0}$. To proceed with the analysis, we need the time derivative of the Hubble parameter, which can be written in terms of the redshift as follows
\begin{align}
    \dot{H} &= -(1+z)H(z)\frac{\mathrm{d} H(z)}{\mathrm{d} z} \nonumber \\
            &= -(1+z)H_{0}^{2}(1+\alpha z + \beta z^2)(\alpha + 2 \beta z).
\end{align}
``It is worth emphasizing that, once the parametric form of $H(z)$ in Eq.~(19) is adopted, no numerical integration of the modified Friedmann equations (11)--(12) is required to evaluate the cosmological quantities of interest. Since the parametrization provides both $H(z)$ and its time derivative $\dot{H}(z)$ in closed analytical form (Eq.(~\ref{parametricH}), the effective energy density $\rho_g$, pressure $p_g$, and equation-of-state parameter $\omega_g$ (Eqs.~(\ref{rhog}, \ref{pg}, \ref{omegag}) can be computed algebraically at any redshift, directly from the best-fit values of $H_0$, $\alpha$, and $\beta$, together with the torsion scalar $T = -6H^2$ (Eq.~(\ref{torsion_scalar})). This model-independent (cosmographic) strategy --- constraining the background kinematics directly from observational data, rather than assuming a matter content and numerically solving the field equations for a fixed $f(T)$ --- has been adopted in similar contexts in modified gravity, such as in $f(Q)$ cosmology \cite{MYRZAKULOV2023101268} and in general parametrizations of scalar-field dark energy \cite{Roy:2022fif}. We note that this approach decouples the observational determination of $H_0, \alpha, \beta$ from the free parameters of $f(T)$ itself ($\Lambda, n, \lambda$), which are not independently constrained by the data within this framework; the resulting $\rho_g, p_g$, and $\omega_g$ profiles (Figs.~\ref{fig:rhoplot}, \ref{fig:pplot}, \ref{fig:oplot}) should therefore be understood as characterizing the qualitative viability of the model given the observationally-constrained kinematics, rather than as a full joint statistical inference over all model parameters.''

The deceleration parameter $q(z)$ is directly derived from the Hubble parameter~\cite{MYRZAKULOV2023101268}:
\begin{equation}
\label{deceleration}
    q(z) = -1 + \frac{(1+z)}{H(z)}\frac{\mathrm{d} H(z)}{\mathrm{d} z} = -1 + \frac{(1+z)(\alpha + 2\beta z)}{1 + \alpha z + \beta z^2}.
\end{equation}

The main objective of this formulation is to test the validity of the proposed parametrization for $H(z)$ against current cosmological observational data, in order to infer realistic values for the constants $\alpha$, $\beta$, and $H_{0}$, and to assess the behavior and viability of the $f(T)$ model previously introduced. For the purpose of analysis, we first tested whether our parametric $H(z)$, using the best-fit values for the constants, can reproduce a universe undergoing accelerated expansion. Afterward, if our $H(z)$ parametrization is valid, we can use the best-fit values to plot the behavior of $\rho_g$, $p_g$, and $\omega_g$.

\section{Testing the parametric $H(z)$}

\label{sec5}

\subsection{The method}
As mentioned previously, the method to obtain the best fits for the parameters $\alpha$, $\beta$, and $H_0$ of $H(z)$ will be based on the maximization of likelihood function. This technique stems from the standard Bayesian approach \cite{LAMPINEN2001257}. The likelihood is defined as
\begin{equation}
\mathcal{L}(\Theta) \propto \exp\left(-\frac{1}{2} \chi^2(\Theta)\right),
\end{equation}
where the chi-squared statistic is given by
\begin{equation}
\label{chi_geral}
\chi^{2}(\Theta) = \sum_{n}\frac{\left[A(z, { \Theta }) - A_{\text{obs}}(z)\right]^2}{\sigma_{A_{\text{obs}}}^2}.
\end{equation}
Let $A$ be an observable quantity at a given redshift $z$, which depends on a set of free parameters $\Theta$. The goal of the analysis is to determine, from $n$ independent measurements, the values that the parameter set $\Theta$ should assume so that the theoretical predictions for $A(z)$ are most likely, given the observed data. In the specific case considered in this work, $\Theta$ corresponds to the parameters of the Hubble function $H(z)$.

\subsection{Cosmic Chronometers}

For the observational dataset, we employ 31 model-independent determinations of the Hubble expansion rate $H(z)$ in the redshift range $0.07 \leq z \leq 1.965$~\cite{michele:01}, derived from the cosmic chronometer technique based on the differential age evolution of passively evolving galaxies. The cosmic chronometer method provides a direct estimate of the expansion rate through the relation between the differential age of the Universe and redshift, without requiring a specific cosmological model for the expansion history~\cite{michele:01}.

To properly account for the statistical and systematic uncertainties affecting the measurements, we construct the covariance matrix following the prescription described in~\cite{michele:01}. The total covariance matrix is written as the sum of the statistical and systematic contributions,

\begin{equation}
\mathrm{Cov}_{ij}=\mathrm{Cov}^{\mathrm{stat}}_{ij}+\mathrm{Cov}^{\mathrm{syst}}_{ij},
\end{equation}

where the systematic contribution is decomposed as

\begin{equation}
\mathrm{Cov}^{\mathrm{syst}}_{ij}=\mathrm{Cov}^{\mathrm{met}}_{ij}+\mathrm{Cov}^{\mathrm{young}}_{ij}+\mathrm{Cov}^{\mathrm{model}}_{ij},
\end{equation}

with the model-dependent contribution further decomposed into

\begin{equation}
\mathrm{Cov}^{\mathrm{model}}_{ij}=\mathrm{Cov}^{\mathrm{SFH}}_{ij}+\mathrm{Cov}^{\mathrm{IMF}}_{ij}+\mathrm{Cov}^{\mathrm{st.lib.}}_{ij}+\mathrm{Cov}^{\mathrm{SPS}}_{ij}.
\end{equation}

Following the prescription of~\cite{michele:01}, the contributions associated with the stellar metallicity, star-formation history (SFH), and possible contamination by a young stellar population are considered uncorrelated between different redshift bins and therefore contribute only to the diagonal elements of the covariance matrix. In contrast, the uncertainties associated with the stellar population synthesis (SPS) modeling are correlated among the measurements at different redshifts and are conservatively treated as fully correlated. The statistical covariance is therefore given by the individual statistical uncertainties, while the systematic contributions are propagated into the diagonal and off-diagonal elements according to their respective correlation properties~\cite{michele:01}. In the published cosmic chronometer datasets, the contributions from metallicity, SFH, and young stellar populations are already included in the quoted measurement uncertainties, while the remaining systematic terms are incorporated through the full covariance prescription~\cite{michele:01}.

The likelihood contribution from the cosmic chronometer measurements is then expressed in matrix form as

\begin{equation}
\chi^{2}_{H}=
\Delta\mathbf{H}^{T}
\mathrm{Cov}^{-1}
\Delta\mathbf{H},
\label{chi2_cc}
\end{equation}

where the residual vector is defined by

\begin{equation}
\Delta H_i =
H(z_i,\alpha,\beta,H_0)-H_{\mathrm{obs}}(z_i),
\end{equation}

and $H(z,\alpha,\beta,H_0)$ is given by Eq.~\eqref{parametricH}. The vector $H_{\mathrm{obs}}$ contains the observational measurements of $H(z)$, while $\mathrm{Cov}$ denotes the full $31\times31$ covariance matrix constructed from the statistical and systematic uncertainties of the cosmic chronometer dataset. This treatment allows the correlations between measurements arising from common systematic effects to be consistently propagated into the cosmological parameter constraints.

\begin{center}
    \vspace{0.5cm}
    \includegraphics[width=0.95\textwidth]{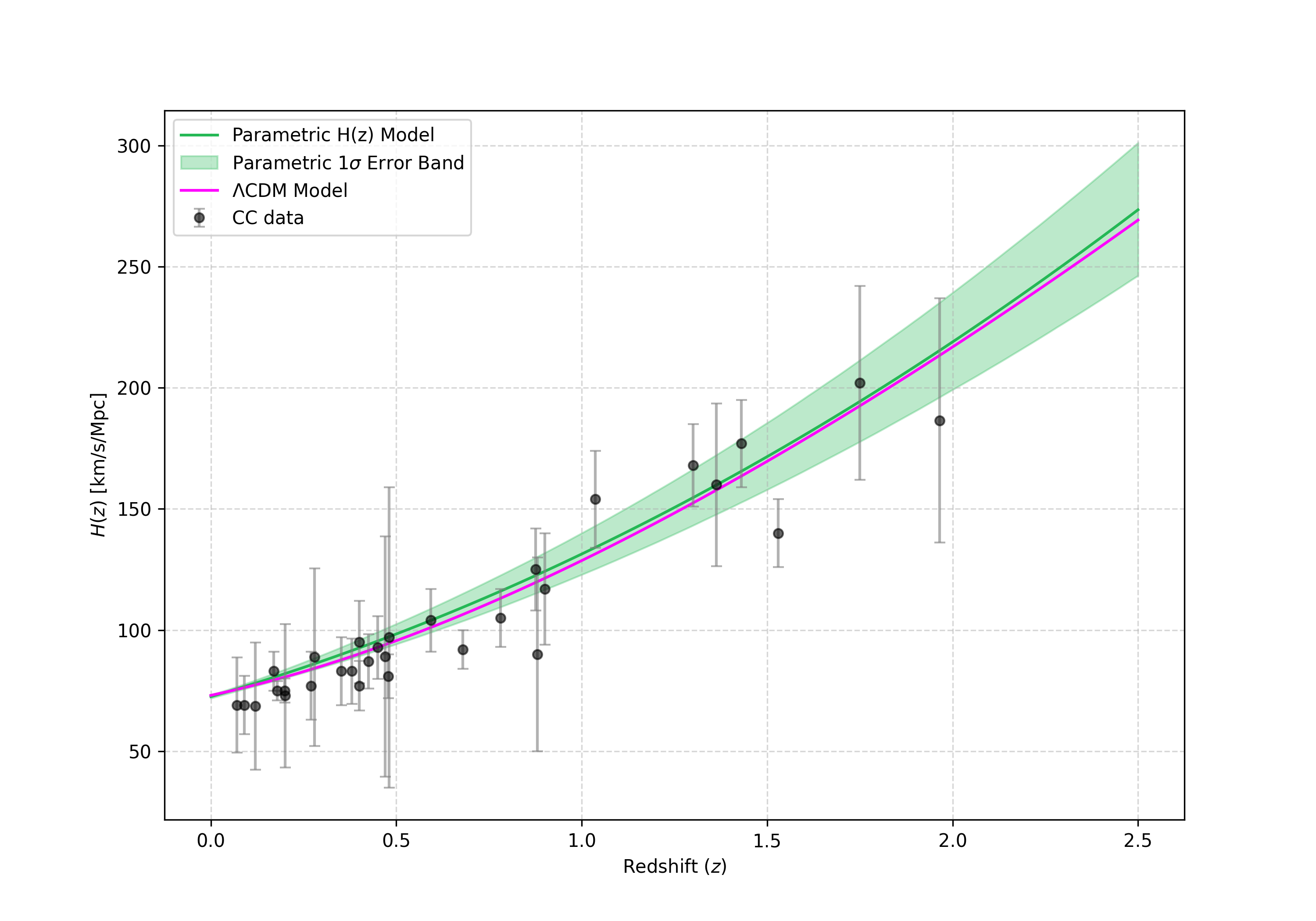}
    \captionof{figure}{Comparison between the two $H(z)$ parametrizations, the parametric $H(z)$ model and the $\Lambda$CDM model (the latter representing a simplified universe containing only dust-like matter). Both curves were computed using the best-fit values reported in Table~\ref{tb2}, and are shown together with the cosmic chronometer Hubble measurements, represented by the black bars.}
    \label{hzdata}
    \vspace{0.5cm}
\end{center}

\subsection{Supernova type Ia dataset}
Commonly called standard candles, Type Ia supernovae are fundamental tools in observational cosmology. Consequently, several collaborative efforts have compiled supernova samples, such as the SuperNova Legacy Survey (SNLS), the Hubble Space Telescope (HST) Survey, and the Sloan Digital Sky Survey (SDSS). One of the most updated datasets is Pantheon+SH0ES \cite{Scolnic_2022}, released in 2022, which catalogs 1701 independent observations covering a redshift range of $0.01 < z < 2.3$. 

Following Ref. \cite{Perivolaropoulos:2023iqj}, the goodness-of-fit parameter $\chi^2$ for the Pantheon+ dataset is constructed as:
\begin{equation}
\label{chi_pan}
    \chi^{2}_{\text{Pantheon+}} = \Delta X^{T} C^{-1}_{\text{stat+sys}} \Delta X = \sum_{i,j=1}^{1701} \Delta X_{i} \left(C^{-1}_{\text{stat+sys}}\right)_{ij} \Delta X_{j},
\end{equation}
where $C^{-1}_{\text{stat+sys}}$ denotes the inverse of the full covariance matrix combining both statistical and systematic uncertainties \cite{Scolnic_2022,MYRZAKULOV2023101268}. The residual vector $\Delta X$ is defined directly as the difference between the observed distance modulus $\mu_{\text{obs}}$ and the theoretical distance modulus prediction $\mu_{\text{model}}(z, \alpha, \beta, H_0)$:
\begin{equation}
    \Delta X_{i} = \mu_{\text{obs}, i} - \mu_{\text{model}}(z_i, \alpha, \beta, H_0),
\end{equation}
which incorporates the calibrated distance modulus data provided by cosmic measurements \cite{Perivolaropoulos:2021jda, Perivolaropoulos:2023iqj}.

The theoretical distance modulus $\mu_{\text{model}}$ is related to the luminosity distance $d_L(z, \alpha, \beta, H_0)$ in megaparsecs ($\text{Mpc}$) by \cite{Perivolaropoulos:2023iqj}:
\begin{equation}
\label{mu_sn}
    \mu_{\text{model}}(z, \alpha, \beta, H_0) = 5 \log_{10} \left( \frac{d_{L}(z, \alpha, \beta, H_0)}{\text{Mpc}} \right) + 25,
\end{equation}
where $d_L$ for a flat FLRW universe is expressed as \cite{Perivolaropoulos:2021jda}:
\begin{equation}
\label{dL_eq}
    d_{L}(z, \alpha, \beta, H_0) = c (1+z) \int^{z}_{0} \frac{dz'}{H(z', \alpha, \beta, H_0)}.
\end{equation}

\begin{center}
    \vspace{0.5cm}
    \includegraphics[width=0.95\textwidth]{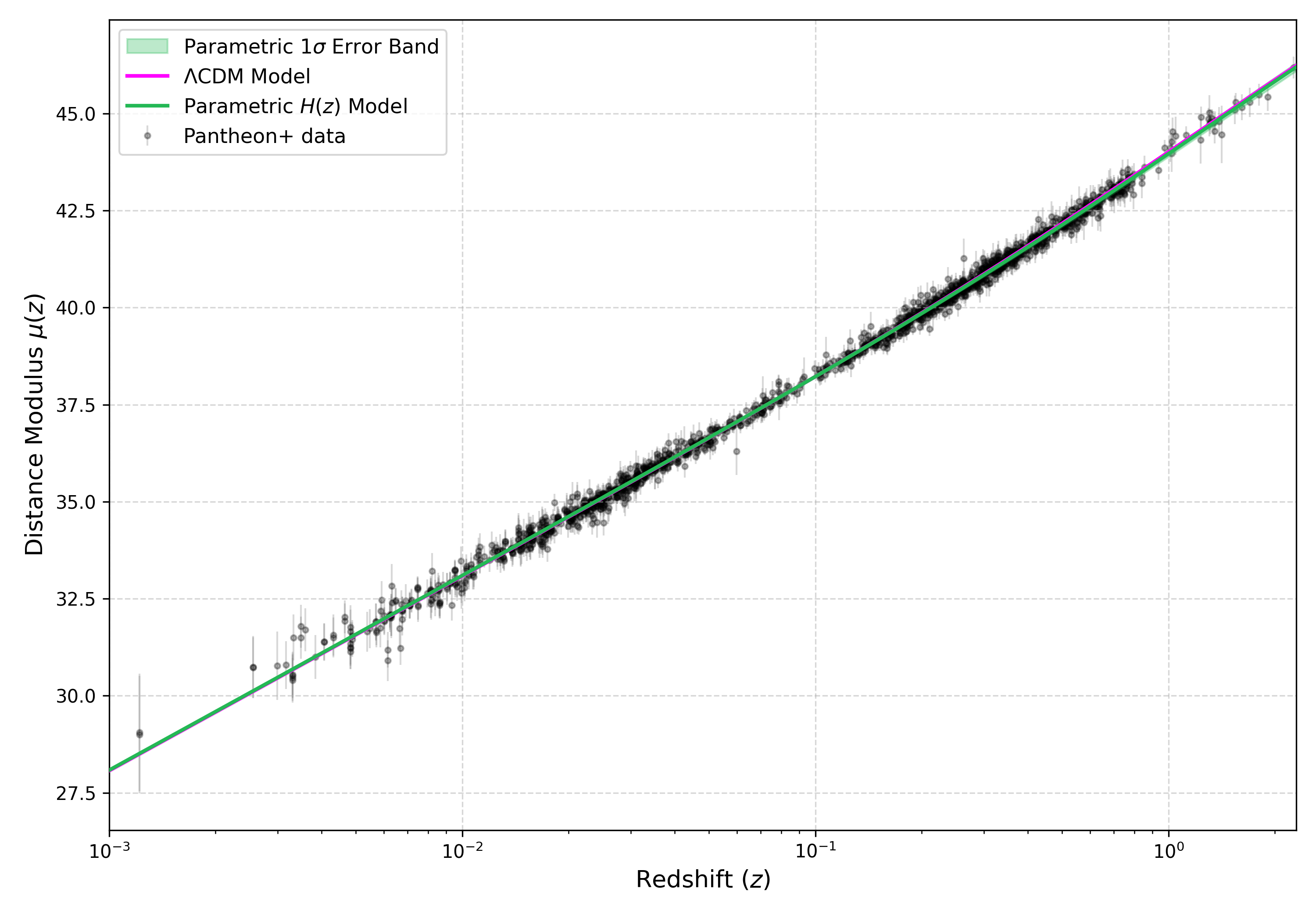}
    \captionof{figure}{Comparison between the two $H(z)$ parametrizations, the parametric $H(z)$ model and the $\Lambda$CDM model, used to construct the apparent magnitude $\mu(z)$ represented in Eq.~\eqref{mu_sn}. Both curves were computed using the best-fit values reported in Table~\ref{tb2}, and are shown together with the Pantheon+SH0ES distance modulus measurements, represented by the black bars.}
    \label{pndata}
    \vspace{0.5cm}
\end{center}

\subsection{Baryon Acoustic Oscillations}

Baryon Acoustic Oscillations (BAO) provide a robust standard ruler to probe the expansion history of the Universe. In this work, we use the second-year Data Release (DR2) from the Dark Energy Spectroscopic Instrument (DESI) \cite{DESI:2025zgx}, which offers high-precision measurements of BAO feature across multiple cosmic tracers covering the redshift range $0.295 < z < 2.33$.

The physical scale of BAO measurements is calibrated by the sound horizon at the drag epoch, $r_d$. Depending on the tracer and the signal-to-noise ratio, DESI provides measurements either as isotropic volume-averaged distances, $D_V(z)/r_d$, or as joint anisotropic measurements of the comoving angular distance, $D_M(z)/r_d$, and the Hubble distance, $D_H(z)/r_d$. The theoretical predictions for these quantities scaled by $r_d$ are defined as follows:
\begin{align}
    \frac{D_H(z)}{r_d} &= \frac{c}{H(z) \, r_d}, \label{eq:DH_rd} \\
    \frac{D_M(z)}{r_d} &= \frac{c}{r_d} \int_{0}^{z} \frac{dz'}{H(z')}, \label{eq:DM_rd} \\
    \frac{D_V(z)}{r_d} &= \left[ z \left( \frac{D_M(z)}{r_d} \right)^2 \left( \frac{D_H(z)}{r_d} \right) \right]^{1/3}, \label{eq:DV_rd}
\end{align}
where $c$ is the speed of light, and $r_d$ is treated as an independent free parameter in our MCMC sampling, making our analysis model-independent of early-Universe cosmic microwave background (CMB) constraints.

The DESI DR2 \cite{DESI:2025zgx} consists of a 1D isotropic measurement ($D_V/r_d$) for the Bright Galaxy Sample (BGS), and 2D correlated anisotropic measurements ($(D_M/r_d, \allowbreak D_H/r_d)$) for Luminous Red Galaxies (LRGs), Emission Line Galaxies (ELGs), the combined LRG+ELG sample, Quasars (QSO), and the Lyman-$\alpha$ (Ly$\alpha$) forest, as summarized in Table \ref{tb1}.

\begin{table*}[t]
\centering
\caption{DESI BAO measurements DR2 \cite{DESI:2025zgx}, including 1D isotropic ($D_V/r_{\mathrm{d}}$) and 2D anisotropic ($D_M/r_{\mathrm{d}}, D_H/r_{\mathrm{d}}$) observables with their correlation coefficient $r_{\mathrm{M,H}}$.}
\label{tb1}
\begin{tabular}{lccccc}
\toprule
Tracer & $z_{\mathrm{eff}}$ & $D_V/r_{\mathrm{d}}$ & $D_M/r_{\mathrm{d}}$ & $D_H/r_{\mathrm{d}}$ & $r_{\mathrm{M,H}}$ \\
\midrule
BGS         & 0.295 & $7.942 \pm 0.075$  & ---                & ---                & ---      \\
LRG1        & 0.510 & ---                & $13.588 \pm 0.167$ & $21.863 \pm 0.425$ & $-0.459$ \\
LRG2        & 0.706 & ---                & $17.351 \pm 0.177$ & $19.455 \pm 0.330$ & $-0.404$ \\
LRG3 + ELG1 & 0.934 & ---                & $21.576 \pm 0.152$ & $17.641 \pm 0.193$ & $-0.416$ \\
ELG2        & 1.321 & ---                & $27.601 \pm 0.318$ & $14.176 \pm 0.221$ & $-0.434$ \\
QSO         & 1.484 & ---                & $30.512 \pm 0.760$ & $12.817 \pm 0.516$ & $-0.500$ \\
Lya         & 2.330 & ---                & $38.988 \pm 0.531$ & $8.632 \pm 0.101$  & $-0.431$ \\
\midrule
LRG3        & 0.922 & ---                & $21.648 \pm 0.178$ & $17.577 \pm 0.213$ & $-0.406$ \\
ELG1        & 0.955 & ---                & $21.707 \pm 0.335$ & $17.803 \pm 0.297$ & $-0.462$ \\
\bottomrule
\end{tabular}
\end{table*}

The total chi-square for the BAO dataset is given by the sum of the 1D and 2D contributions:
\begin{equation}
    \chi^2_{\rm BAO} = \chi^2_{\rm 1D} + \chi^2_{\rm 2D}.
\end{equation}
For the 1D data points, the chi-square contribution is simple:
\begin{equation}
\label{chi_bao1}
    \chi^2_{\rm 1D} = \sum_{i=1}^{2} \left( \frac{\left[D_V(z_i)/r_d\right]_{\rm obs} - \left[D_V(z_i)/r_d\right]_{\rm pred}}{\sigma_{D_V, i}} \right)^2.
\end{equation}
For the 2D correlated data points, the chi-square contribution is computed using the $2 \times 2$ covariance matrix $\mathbf{C}_j$ for each redshift bin $j$:
\begin{equation}
\label{chi_bao2}
    \chi^2_{\rm 2D} = \sum_{j=1}^{5} \mathbf{\Delta}_j^T \, \mathbf{C}_j^{-1} \, \mathbf{\Delta}_j,
\end{equation}
where $\mathbf{\Delta}_j$ is the residual vector between observed and predicted values:
\begin{equation}
    \mathbf{\Delta}_j = \begin{pmatrix} 
    (D_M/r_d)_{\rm obs} - (D_M/r_d)_{\rm pred} \\ 
    (D_H/r_d)_{\rm obs} - (D_H/r_d)_{\rm pred} 
    \end{pmatrix}_j,
\end{equation}
and the covariance matrix $\mathbf{C}_j$ is constructed as:
\begin{equation}
    \mathbf{C}_j = \begin{pmatrix} 
    \sigma_{D_M}^2 & r \, \sigma_{D_M} \sigma_{D_H} \\ 
    r \, \sigma_{D_M} \sigma_{D_H} & \sigma_{D_H}^2 
    \end{pmatrix}_j.
\end{equation}

\begin{center}
    \vspace{0.5cm}
    \includegraphics[width= 0.95\textwidth]{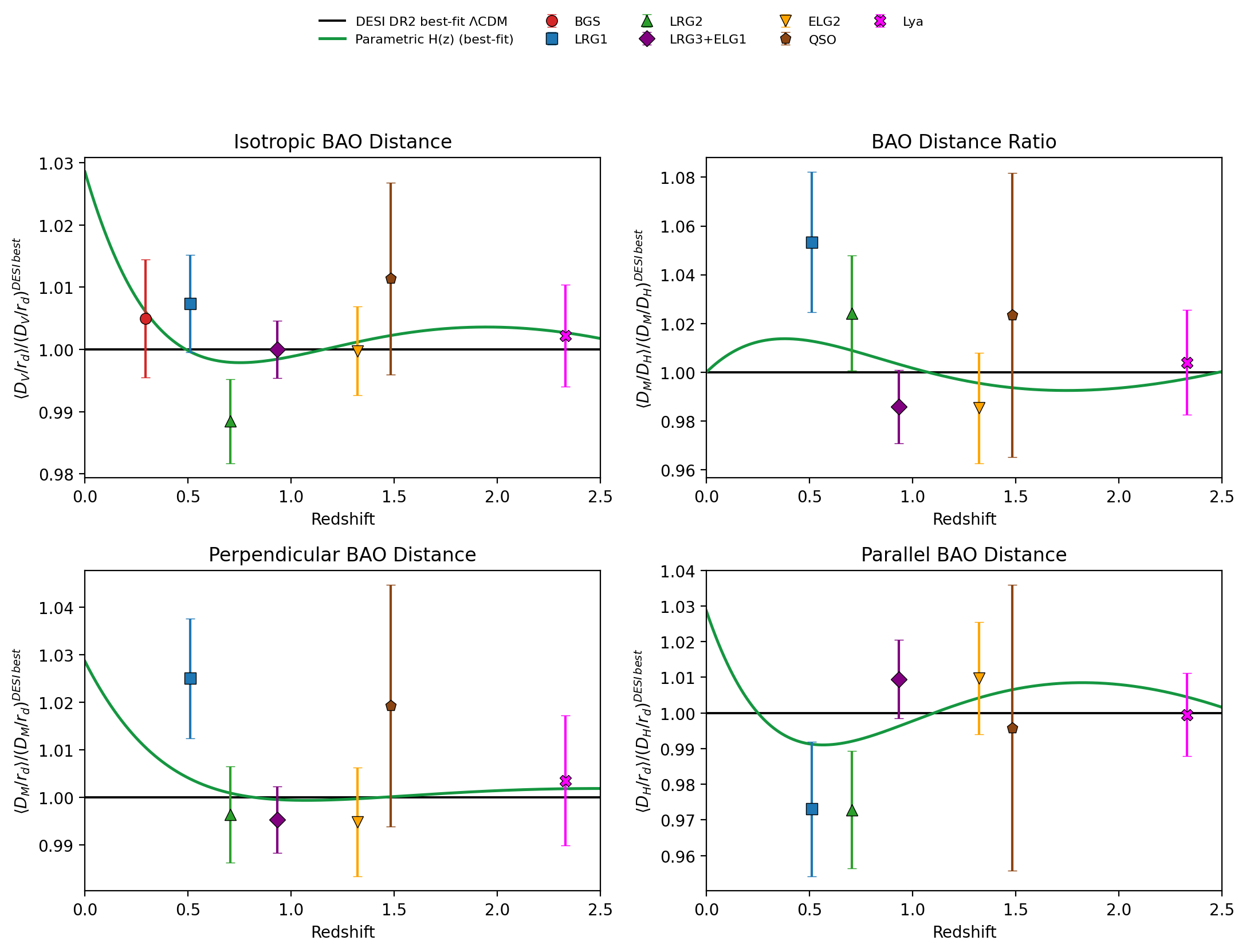}
    \captionof{figure}{Fractional residuals of the DESI DR2 BAO measurements relative to the best-fit $\Lambda$CDM model, shown for the isotropic volume-averaged distance $D_V/r_d$ (top left), the distance ratio $D_M/r_d \, / \, D_H/r_d$ (top right), the comoving angular distance $D_M/r_d$ (bottom left), and the Hubble distance $D_H/r_d$ (bottom right), as a function of redshift. The black horizontal line marks the DESI DR2 best-fit $\Lambda$CDM prediction (ratio $=1$), and the colored markers show the individual BAO tracers (BGS, LRG1, LRG2, LRG3+ELG1, ELG2, QSO, and Ly$\alpha$) with their $1\sigma$ error bars. The green curve shows the prediction of the parametric $H(z)$ model using the best-fit values of $H_0$, $\alpha$, $\beta$, and $r_d$ best fits presents in Table~\ref{tb2}, normalized to the same $\Lambda$CDM reference.}
    \label{desidata}
    \vspace{0.5cm}
\end{center}

\section{Results}
\label{sec6}

\begin{figure*}
    \centering
    \includegraphics[scale=0.5]{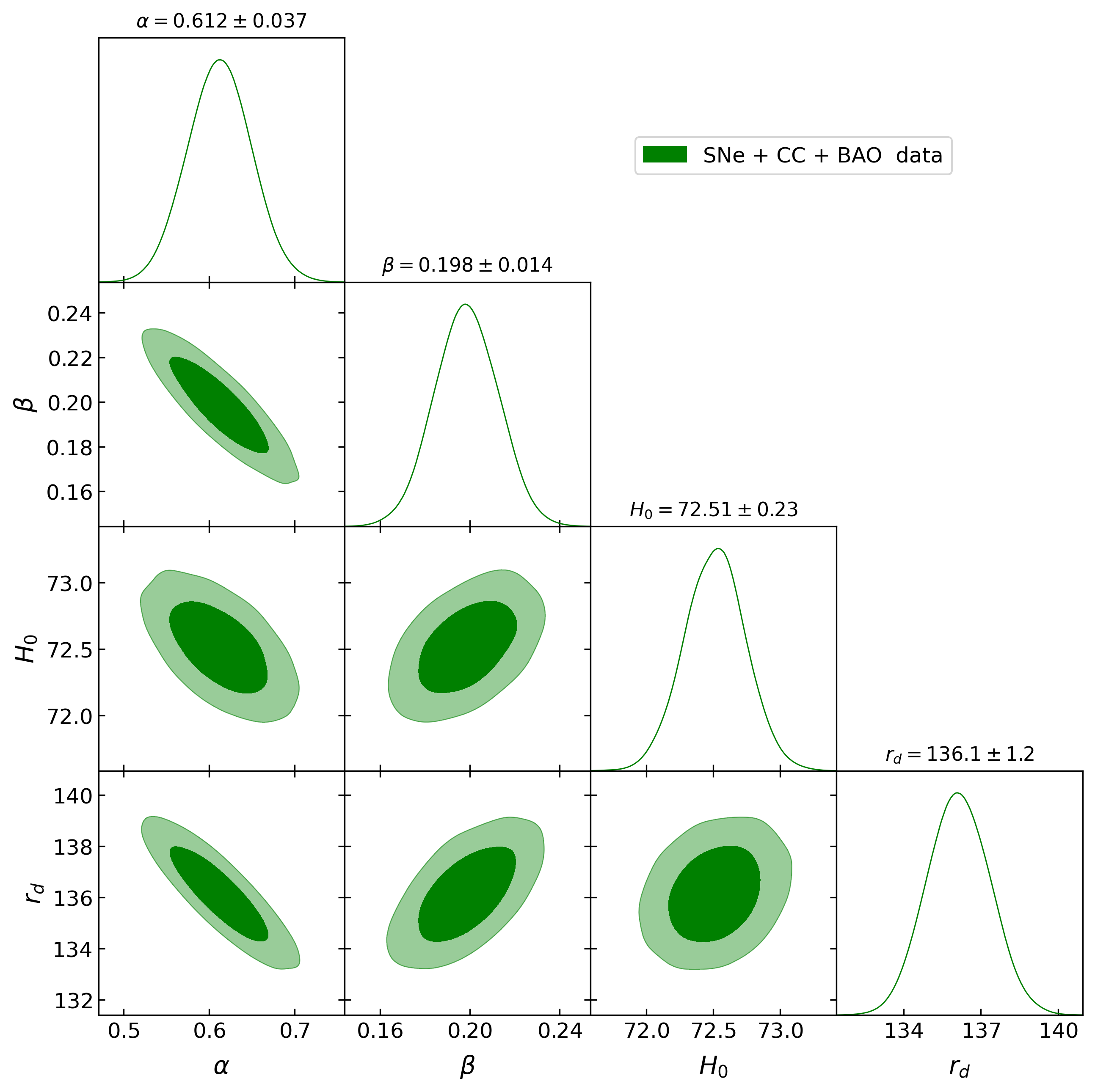}
    \caption{Contours of the confidence regions of the quadratic expansion model $H(z)$ tested for data from Pantheon+, BAO, and Hubble datasets in the maximization of likelihood method. The regions from outside to inside represent, respectively, $1\sigma$ and $2\sigma$. The marginalized $1\sigma$ values displayed above each 1D panel were computed via \texttt{GetDist}'s kernel density estimation and are reported here as symmetric uncertainties for visualization purposes; they may therefore differ slightly from the asymmetric $68\%$ credible intervals obtained directly from the raw MCMC percentiles and quoted in Table~\ref{tb2}.}
    \label{triplot}
\end{figure*}

In this section, we will discuss the evolution of the parameters of the model $f(T)$ given by Eq.\eqref{lag1}. It was previously mentioned about the possibility of considering extra terms (in relation to those already existing in $\Lambda CDM$) such as an effective energy density $\rho_{g}$ and a pressure $p_{g}$ of geometrical origin in $f(T)$ gravity, and now with the numerical values of $H_0$, $\alpha$ and $\beta$ defined by observational data, it becomes possible to perform an analysis of the viability and behavior of the model $f(T)$ in the scenario where the quadratic expansion of $H(z)$ successfully reproduces a universe undergoing accelerated expansion. In this regime, the best-fit parameters remain consistent with observational constraints in Table \ref{tb2}, ensuring that the model provides a physically acceptable description of late-time cosmic acceleration.
After submitting the $H(z)$ model to obtain the parameters $H_0$, $\alpha$ and $\beta$ by the maximization of likelihood method given by equations \eqref{chi2_cc}, \eqref{chi_pan}, \eqref{chi_bao1} and \eqref{chi_bao2}, and with the support of the emcee Python library to perform the MCMC sampling \cite{foreman2013emcee}, for the 1701 (Fig.~\ref{pndata}) data points of Ia supernovae, the 31 data points of the Hubble dataset (Fig.~\ref{hzdata}) and the 9 BAO tracers from DESI DR2 (Table~\ref{tb1}), comprising 1 isotropic ($D_V/r_d$) and 8 anisotropic measurements (Fig.~\ref{desidata}) (each contributing $D_M/r_d$ and $D_H/r_d$), for a total of 9 BAO observables, we obtained results within the expected range. For comparison purposes, we also tested the simplified flat $\Lambda CDM$ model containing dust-type matter and radiation by $ 1 + z_{eq} = \frac{\Omega_r}{\Omega_m}$ formula, for the same data. The combination of datasets occurred by the product of the individual likelihoods $\mathcal{L}$, so we have that $\mathcal{L}_{\text{total}} = 
\mathcal{L}_{H} \cdot 
\mathcal{L}_{\text{Pantheon+}} \cdot 
\mathcal{L}_{\text{BAO}}$. The best fits found for the quadratic expansion of the parametric $H(z)$ model and the flat $\Lambda CDM $ model for the same data are presented in Table \ref{tb2}, and the confidence regions for parametric $H(z)$ model results are represented in Fig.\ref{triplot}, plotted with the Getdist Python Library \cite{Lewis:2019xzd}.

\begin{table*}[htbp]
\caption{Best-fit parameters and $\chi^{2}$ values for the parametric $H(z)$ model and the $\Lambda$CDM model with $1\sigma$ uncertainties.}
\label{tb2}
\centering
\begin{tabular}{l c c c}
\hline
Model & Parameter & Best-fit value & $\chi^2_{\min}$ \\
\hline
Parametric function $H(z)$ & $H_{0}$ & $72.518^{+0.224}_{-0.230}$ & \multirow{4}{*}{1078.811} \\
 & $\alpha$ & $0.612^{+0.038}_{-0.038}$ & \\
 & $\beta$ & $0.198^{+0.015}_{-0.014}$ & \\
 & $r_d$ & $136.125^{+1.263}_{-1.234}$ & \\
\hline
$\Lambda$CDM & $H_{0}$ & $73.017^{+0.198}_{-0.198}$ & \multirow{3}{*}{1087.816} \\
 & $\Omega_{m}$ & $0.301^{+0.008}_{-0.008}$ & \\
 & $r_d$ & $138.634^{+0.920}_{-0.928}$ & \\
\hline
\end{tabular}
\end{table*}

\begin{figure*}
    \centering
    \includegraphics[scale = 0.5]{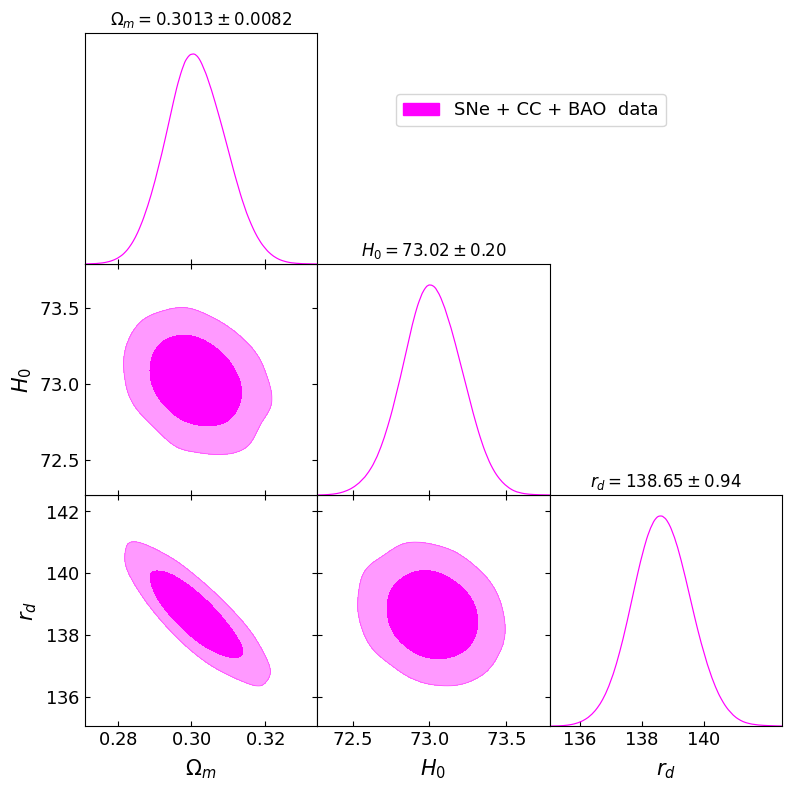}
    \caption{Contours of the confidence regions of $\Lambda$CDM tested for data from Pantheon+, BAO, and Hubble datasets in the maximization of likelihood method. The regions from outside to inside represent, respectively, $1\sigma$ and $2\sigma$. The marginalized $1\sigma$ values displayed above each 1D panel were computed via \texttt{GetDist}'s kernel density estimation and are reported here as symmetric uncertainties for visualization purposes; they may therefore differ slightly from the (in this case nearly symmetric) $68\%$ credible intervals obtained directly from the raw MCMC percentiles and quoted in Table~\ref{tb2}.}
    \label{triplot_lcdm}
\end{figure*}
The next step is to verify whether the parametrization of $H(z)$ reproduces an accelerated universe, and for this, the deceleration parameter $q$ must be negative for acceleration. Using Eq.~\eqref{deceleration} and the best-fit values from Table \ref{tb2}, Fig.~\ref{qplot} shows the behavior of $q(z)$. Both models do not exhibit constant acceleration, as $q$ becomes positive for $z \gtrsim 1$, indicating deceleration. However, the recent universe remains in accelerated expansion, with $q_0^{\Lambda\mathrm{CDM}} = -0.5477^{+0.0379}_{-0.0354}$ and $q_0^{H(z)} = -0.388_{-0.038}^{+0.038}$ in the limit $z \to 0$, at the $1\sigma$ confidence level.

\begin{center}
    \vspace{0.3cm}
    \includegraphics[width=0.95\textwidth]{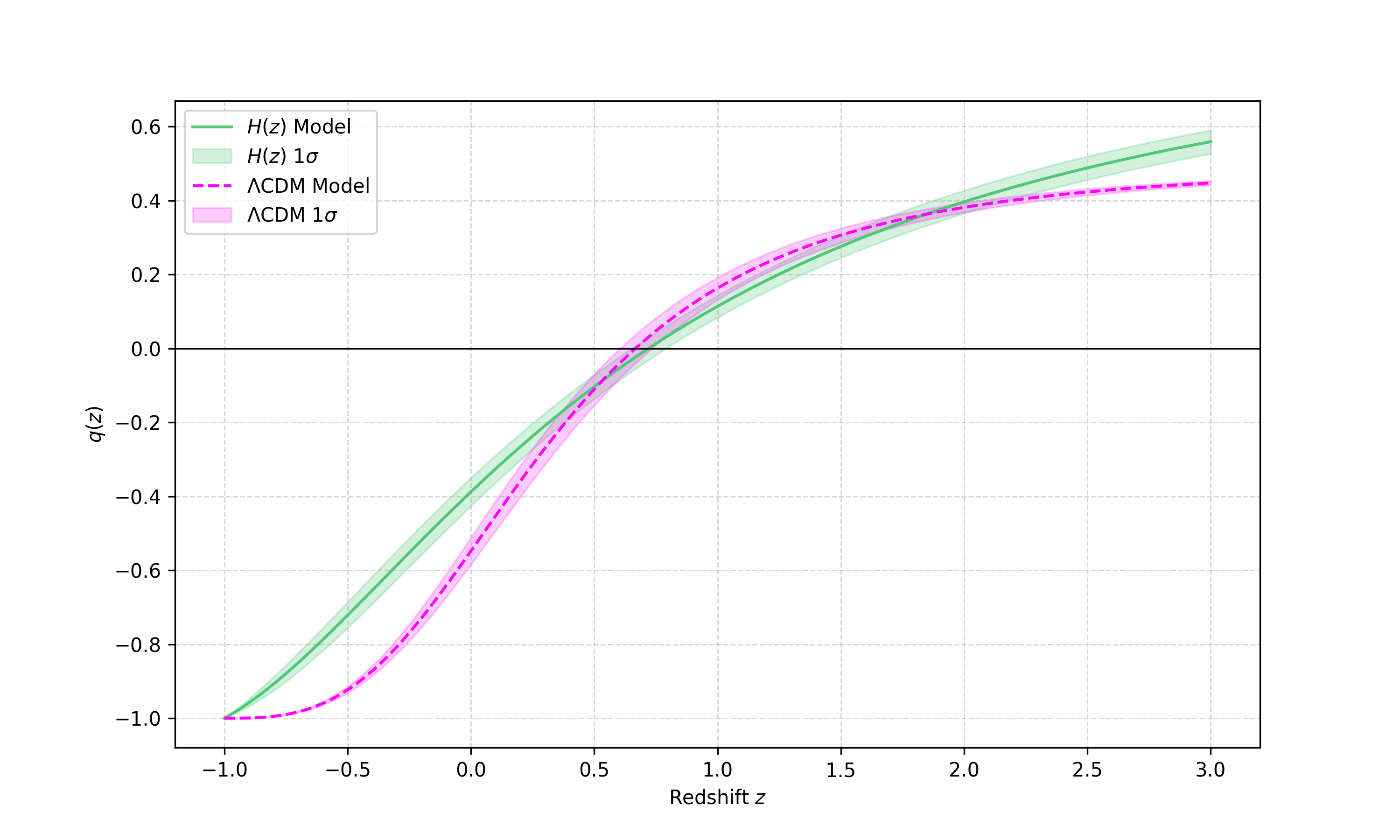}
    \captionof{figure}{Deceleration parameter $q$ for the parameterized $H(z)$ and $\Lambda\text{CDM}$ models. The best-fit values presented in Table~\ref{tb2} for each model were used to generate the plot.}
    \label{qplot}
    \vspace{0.3cm}
\end{center}

\begin{figure*}[htbp]
\centering
  \includegraphics[width=\textwidth]{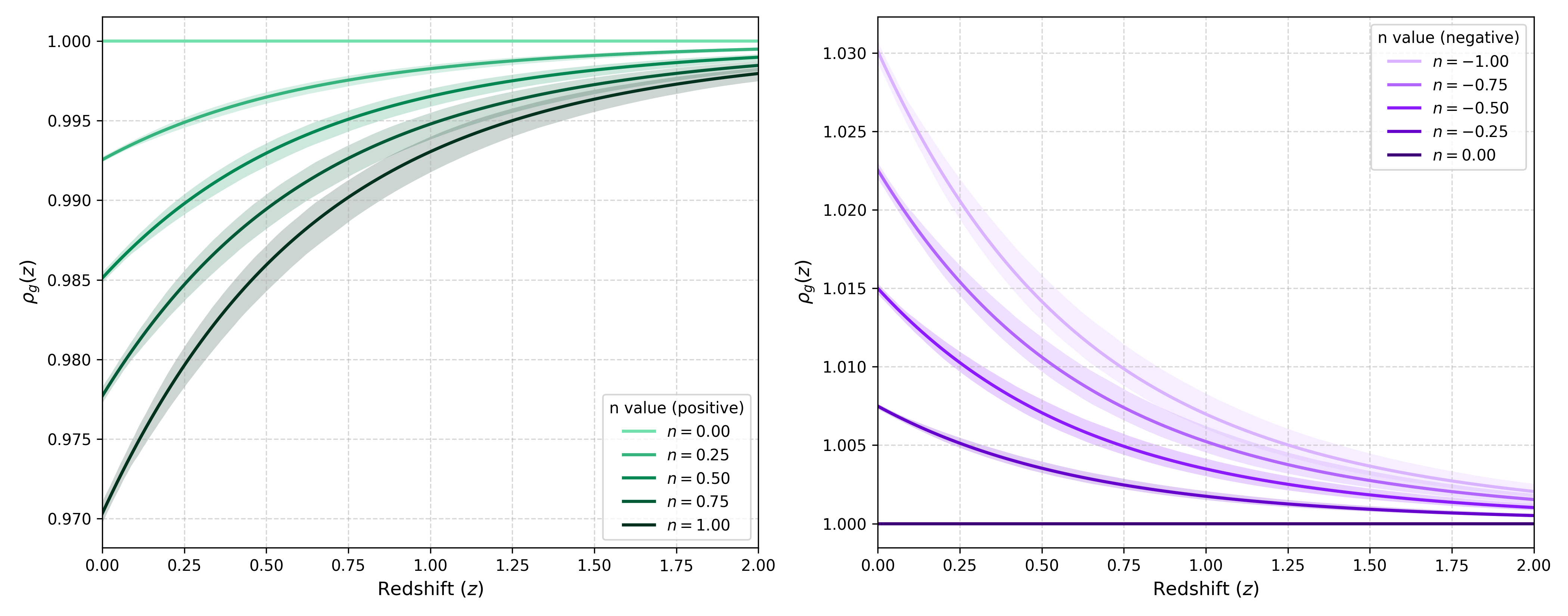}
  \caption{Evolution of the effective energy density $\rho_g(z)$, normalized to $\Lambda=1$, for positive (left panel) and negative (right panel) values of $n$. The curves were constructed using the best-fit values of $H_0$, $\alpha$, and $\beta$ reported in Table~\ref{tb2}. For positive $n$, $\rho_g$ decreases below unity at low redshift, with the deviation from the cosmological-constant behavior ($n=0$) growing monotonically with $n$; for negative $n$, $\rho_g$ instead increases above unity, with larger $|n|$ producing a larger deviation. In both cases, $\rho_g \to 1$ at high redshift, recovering the cosmological-constant limit, while the shaded bands indicate the propagated $1\sigma$ uncertainty from the best-fit parameters. For illustrative purposes, the numerical values $\lambda=10^{-3}$ and $\Lambda=1$ were adopted.}
  \label{fig:rhoplot}
\end{figure*}

For the density $\rho$ and the pressure $p_{g}$, we have to return to equations \eqref{rhog} and \eqref{pg}, applying the function $f(T)$ from Eq.\eqref{lag1} and using Eq.\eqref{torsion_scalar} together with the parameterization of $H(z)$. We also tested the hypothesis of negative values for $n$, which generated results consistent with and ``expected'', with $\rho_{g}$ being positive and $p_{g}$ being negative. However, since we are dealing with something effective and not with a real fluid, the problem is mitigated, at least in terms of interpretation. The graphical representations of $\rho_{g}$ and $p_{g}$ are illustrated in Figs. \ref{fig:rhoplot} and \ref{fig:pplot}, respectively.

\begin{center}
\begin{figure}
\centering
  \includegraphics[width=\textwidth]{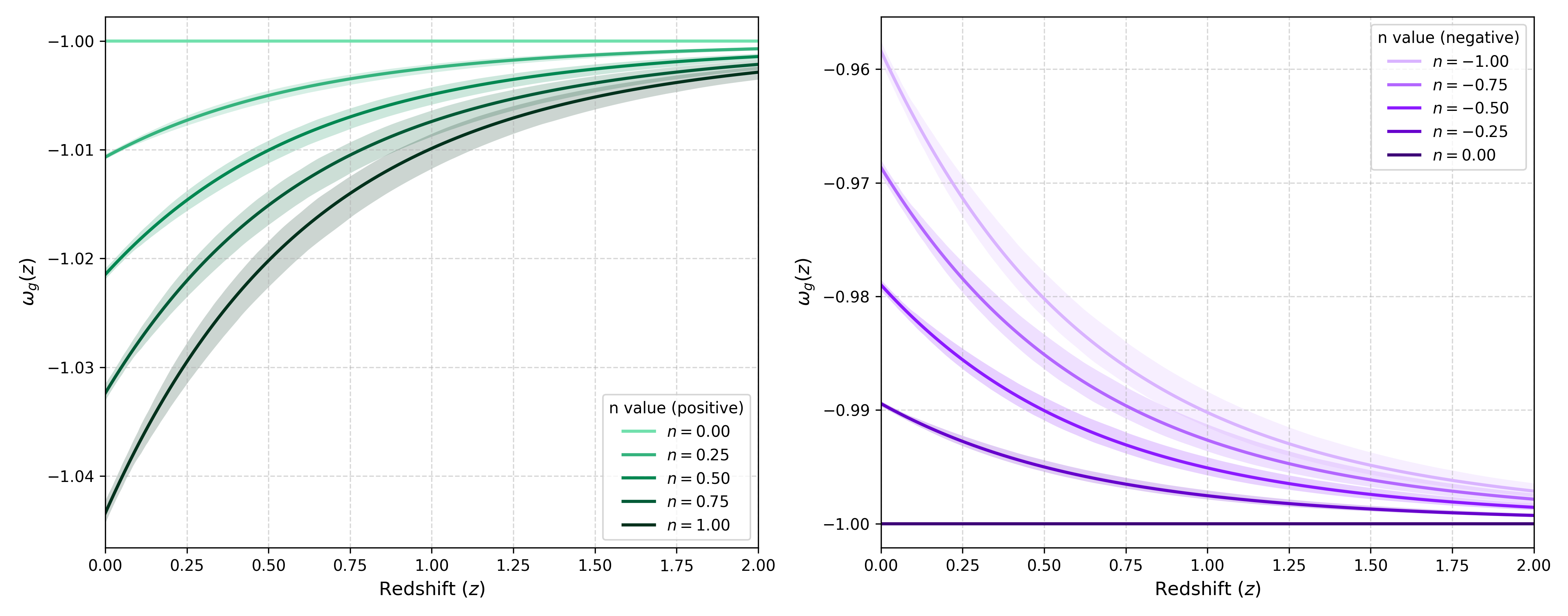}
    \caption{Evolution of the effective equation-of-state parameter $\omega_g(z)$ for positive (left panel) and negative (right panel) values of $n$. The curves were constructed using the best-fit values of $H_0$, $\alpha$, and $\beta$ reported in Table~\ref{tb2}. For positive $n$, $\omega_g$ drops below $-1$ at low redshift, exhibiting phantom-like behavior that grows more pronounced with increasing $n$; for negative $n$, $\omega_g$ instead remains above $-1$, exhibiting quintessence-like behavior that grows more pronounced with increasing $|n|$. In both cases, $\omega_g \to -1$ as $z \to \infty$, recovering the cosmological-constant limit. For illustrative purposes, the numerical values $\lambda=10^{-3}$ and $\Lambda=1$ were adopted.}
    \label{fig:oplot}
\end{figure}
\end{center}

For the state parameter $\omega_{g}$ of the fluid we need to return to Eq.\eqref{omegag} and by using the same applications that we used for $\rho_{g}$ and $p_{g}$, we obtain the effective dynamic form of $\omega$ represented in Fig~\ref{fig:oplot} By taking the limit of $\omega$ for $z$ tending to $\infty$ we have that $\omega = -1$. It is important to mention that the equations for the energy density, pressure, and equation of state parameter are omitted here, as they are too lengthy.

\begin{center}
\centering
  \includegraphics[width=\textwidth]{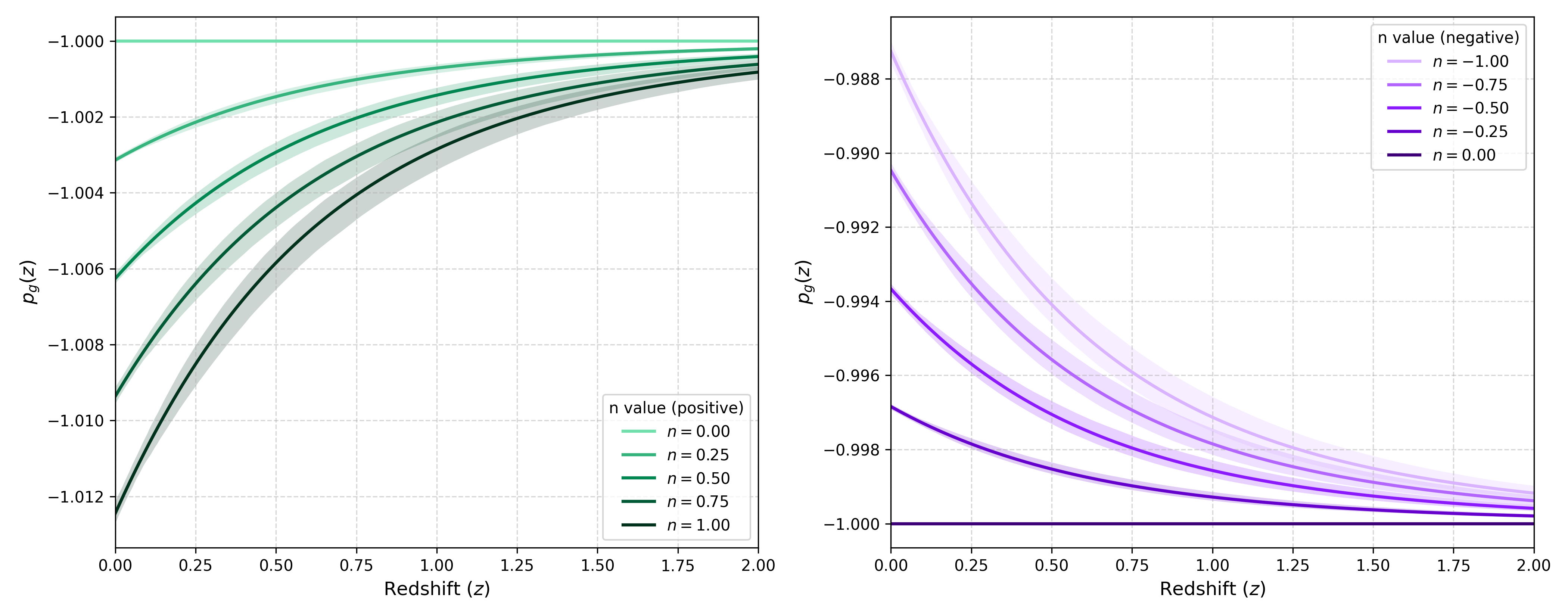}
    \captionof{figure}{Evolution of the effective pressure $p_g(z)$, normalized to $\Lambda=1$, for positive (left panel) and negative (right panel) values of $n$. The curves were constructed using the best-fit values of $H_0$, $\alpha$, and $\beta$ reported in Table~\ref{tb2}. For positive $n$, $p_g$ decreases below the cosmological-constant value ($p_g=-1$, corresponding to $n=0$) at low redshift, with the deviation growing monotonically with $n$; for negative $n$, $p_g$ instead increases above $-1$, with larger $|n|$ producing a larger deviation. In both cases, $p_g \to -1$ at high redshift, recovering the cosmological-constant limit, while the shaded bands indicate the propagated $1\sigma$ uncertainty from the best-fit parameters. For illustrative purposes, the numerical values $\lambda=10^{-3}$ and $\Lambda=1$ were adopted.}
    \label{fig:pplot}
    \vspace{0.3cm}
\end{center}

\vspace{-1mm} 

Thus, within the proposed $f(T)$ model, it is possible to state that, both when referring to the Big Bang and when referring to the distant future, the solution tends to appear to be a cosmological constant. Another interesting factor is the possibility of plausible phantom type solutions for the case in which $n$ assumes negative values, without losing the characteristic of $\omega_{g} = -1$ for limit $\infty$. From such characteristics, despite being a modified gravity model, the behavior of the model regarding the dark energy sector would be the same as a Quintessence or phantom model depending on the sign of $n$. This opens the possibility of there being some cosmological model with modifications in the matter and energy content that can reproduce such results without the need to make modifications in the gravitational sector, such as the $\omega CDM$ models with direct changes in the state parameter of the fluid responsible for cosmic expansion, such as the CPL parameterization of $\omega$ \cite{Chevallier:2000qy,PhysRevLett.90.091301}.

\subsection{Model Selection: AIC and BIC Analysis}

To evaluate whether the additional degree of freedom in our phenomenological $H(z)$ parametrization is statistically supported by the data, we perform a model selection analysis using the Akaike Information Criterion (AIC) and the Bayesian Information Criterion (BIC). For Gaussian errors, these metrics are defined as:
\begin{equation}
\mathrm{AIC} = \chi^2_{\min} + 2k,
\end{equation}
\begin{equation}
\mathrm{BIC} = \chi^2_{\min} + k\ln(N),
\end{equation}
where $\chi^2_{\min}$ is the minimum chi-squared value, $k$ is the number of free parameters, and $N$ is the total size of the dataset considered \cite{Liddle:2004nh,burnham2007model}. We define $\Delta\mathrm{IC} = \mathrm{IC}_{\mathrm{model}} - \mathrm{IC}_{\Lambda\mathrm{CDM}}$, so that a negative value favors the parametric $H(z)$ model.

Our full observational sample consists of $N = 1741$ points (31 Cosmic Chronometers, 9 BAO measurements from DESI DR2, and 1701 points from Pantheon+/SH0ES). Table~\ref{tb3} presents the statistical metrics for both $\Lambda$CDM ($k=2$) and our $H(z)$ template ($k=3$) using the full combined dataset.

\begin{table}[h]
\centering
\caption{Model selection criteria comparing $\Lambda$CDM and our phenomenological $H(z)$ model for $N=1741$ observations (CC+BAO+SNe).}
\label{tb3}
\begin{tabular}{lccccccc}
\hline
Model & $k$ & $\chi^2_{\min}$ & AIC & BIC & $\Delta$AIC & $\Delta$BIC \\
\hline
$\Lambda$CDM   & 2 & 1087.817 & 1091.817 & 1102.741 & 0 & 0 \\
$H(z)$ Model   & 3 & 1078.811 & 1084.811 & 1101.198 & $-7.006$ & $-1.543$ \\
\hline
\end{tabular}
\end{table}

As shown in Table~\ref{tb3}, we obtain $\Delta\mathrm{AIC} = -7.006$, which indicates considerable support for our parametrization over the standard cosmological model, falling in the ``considerably less support'' range for $\Lambda$CDM on the Akaike scale ($4 < |\Delta\mathrm{AIC}| < 7$) and approaching the threshold conventionally associated with strong evidence ($|\Delta\mathrm{AIC}| > 10$) \cite{burnham2007model}. Under the BIC, which imposes a more stringent penalty for the large sample size considered here ($\ln(1741) \approx 7.462$), the difference narrows to $\Delta\mathrm{BIC} = -1.543$. On the Kass--Raftery/Jeffreys scale \cite{Kass:1995loi, Trotta:2008qt}, this falls within the ``weak'' evidence band ($0 < |\Delta\mathrm{BIC}| < 2$), indicating that the improvement in fit ($\Delta\chi^2 = -9.006$) is only partially favored once the extra degree of freedom is penalized more heavily. Taken together, the two criteria point in the same direction, indicating a mild-to-moderate preference for the $H(z)$ parametrization, while disagreeing on strength: the AIC comparison is more favorable than the BIC comparison, a discrepancy expected given BIC's stronger penalty on model complexity for $N=1741$.

To assess the robustness of this result and identify which subset(s) of the data drive the preference for the parametric model, we repeat the same AIC/BIC comparison for the CC, BAO, and SNe datasets individually and in pairwise combination (CC+BAO, CC+SNe, and SNe+BAO). Table~\ref{tb5} presents the results for each dataset analyzed independently, and Table~\ref{tb6} presents the pairwise combinations.

\begin{table}[h]
\centering
\caption{Best-fit parameters, $\chi^2_{\min}$ values, AIC, BIC, $\Delta$AIC, and $\Delta$BIC for the parametric $H(z)$ model and the $\Lambda$CDM model using individual datasets with $1\sigma$ uncertainties.}
\label{tb5}
\resizebox{\textwidth}{!}{%
\begin{tabular}{lccccccccc}
\hline
\multicolumn{10}{c}{Parametric $H(z)$ Model} \\
\hline
Data Sets & $H_0$ [km/s/Mpc] & $\alpha$ & $\beta$ & $r_d$ [Mpc] & $\chi^2_{\min}$ & AIC & BIC & $\Delta$AIC & $\Delta$BIC \\
\hline
CC  & $67.416^{+5.213}_{-4.512}$ & $0.507^{+0.246}_{-0.259}$ & $0.287^{+0.151}_{-0.146}$ & --                            & 14.723   & 20.723   & 25.024   & 2.245    & 3.679 \\
BAO & $70.436^{+6.385}_{-5.949}$ & $0.608^{+0.048}_{-0.046}$ & $0.200^{+0.017}_{-0.017}$ & $140.362^{+12.703}_{-11.621}$ & 7.598    & 15.598   & 16.387   & $-0.953$ & $-0.953$ \\
SNe & $72.378^{+0.314}_{-0.289}$ & $0.649^{+0.081}_{-0.090}$ & $0.216^{+0.265}_{-0.147}$ & --                            & 1052.406 & 1058.406 & 1074.723 & 0.190    & 5.593 \\
\hline
\multicolumn{10}{c}{$\Lambda$CDM Model} \\
\hline
Data Sets & $H_0$ [km/s/Mpc] & $\Omega_{m,0}$ & & $r_d$ [Mpc] & $\chi^2_{\min}$ & AIC & BIC & $\Delta$AIC & $\Delta$BIC \\
\hline
CC  & $67.530^{+4.806}_{-4.484}$ & $0.325^{+0.063}_{-0.057}$ & & --                           & 14.478   & 18.478   & 21.345   & -- & -- \\
BAO & $71.359^{+5.788}_{-5.508}$ & $0.297^{+0.009}_{-0.009}$ & & $142.336^{+11.977}_{-10.760}$ & 10.551   & 16.551   & 17.340   & -- & -- \\
SNe & $72.460^{+0.299}_{-0.296}$ & $0.400^{+0.039}_{-0.041}$ & & --                           & 1054.216 & 1058.216 & 1069.130 & -- & -- \\
\hline
\end{tabular}%
}
\end{table}

For CC, the parametric model is disfavored relative to $\Lambda$CDM under both criteria ($\Delta\mathrm{AIC} = 2.245$, $\Delta\mathrm{BIC} = 3.679$), while for SNe the two criteria diverge, with a negligible $\Delta\mathrm{AIC} = 0.190$ but a substantially larger $\Delta\mathrm{BIC} = 5.593$ reflecting the stronger sample-size penalty imposed by BIC on the larger SNe compilation. Only the BAO dataset shows a mild preference for the parametric model ($\Delta\mathrm{AIC} = \Delta\mathrm{BIC} = -0.953$), though the comparatively small number of BAO data points limits the statistical weight of this result. Taken together, these results indicate that neither CC nor SNe alone contain sufficient information to detect a deviation from $\Lambda$CDM at the level of the quadratic $H(z)$ expansion, which is consistent with the expectation that any genuine signal should only emerge once complementary probes, in particular the geometric distance information encoded in BAO, are incorporated.

\begin{table}[h]
\centering
\caption{Best-fit parameters, $\chi^2_{\min}$ values, AIC, BIC, $\Delta$AIC, and $\Delta$BIC for the parametric $H(z)$ model and the $\Lambda$CDM model using pairwise-combined datasets with $1\sigma$ uncertainties.}
\label{tb6}
\resizebox{\textwidth}{!}{%
\begin{tabular}{lccccccccc}
\hline
\multicolumn{10}{c}{Parametric $H(z)$ Model} \\
\hline
Data Sets & $H_0$ [km/s/Mpc] & $\alpha$ & $\beta$ & $r_d$ [Mpc] & $\chi^2_{\min}$ & AIC & BIC & $\Delta$AIC & $\Delta$BIC \\
\hline
CC + BAO  & $67.407^{+3.351}_{-3.057}$ & $0.607^{+0.047}_{-0.045}$ & $0.201^{+0.016}_{-0.016}$ & $146.559^{+6.773}_{-6.578}$ & 22.598   & 30.598   & 37.354   & $-0.598$ & 1.091 \\
CC + SNe  & $72.392^{+0.311}_{-0.288}$ & $0.654^{+0.070}_{-0.074}$ & $0.134^{+0.076}_{-0.068}$ & --                          & 1070.327 & 1076.327 & 1092.698 & $-2.790$ & 2.667 \\
SNe + BAO & $72.473^{+0.232}_{-0.230}$ & $0.627^{+0.040}_{-0.039}$ & $0.194^{+0.015}_{-0.015}$ & $135.623^{+1.306}_{-1.263}$ & 1061.192 & 1069.192 & 1090.969 & $-8.537$ & $-3.093$ \\
\hline
\multicolumn{10}{c}{$\Lambda$CDM Model} \\
\hline
Data Sets & $H_0$ [km/s/Mpc] & $\Omega_{m,0}$ & & $r_d$ [Mpc] & $\chi^2_{\min}$ & AIC & BIC & $\Delta$AIC & $\Delta$BIC \\
\hline
CC + BAO  & $69.492^{+3.260}_{-3.133}$ & $0.298^{+0.008}_{-0.008}$ & & $146.054^{+6.785}_{-6.451}$ & 25.196   & 31.196   & 36.263   & -- & -- \\
CC + SNe  & $72.788^{+0.256}_{-0.249}$ & $0.338^{+0.028}_{-0.027}$ & & --                          & 1075.117 & 1079.117 & 1090.031 & -- & -- \\
SNe + BAO & $73.020^{+0.200}_{-0.198}$ & $0.302^{+0.009}_{-0.008}$ & & $138.528^{+0.944}_{-0.966}$ & 1071.729 & 1077.729 & 1094.062 & -- & -- \\
\hline
\end{tabular}%
}
\end{table}

For CC+BAO and CC+SNe, the two information criteria disagree in sign ($\Delta\mathrm{AIC} = -0.598$, $\Delta\mathrm{BIC} = +1.091$ for CC+BAO; $\Delta\mathrm{AIC} = -2.790$, $\Delta\mathrm{BIC} = +2.667$ for CC+SNe), indicating that neither combination provides decisive evidence for either model once the extra flexibility introduced by the quadratic expansion is weighed against the modest improvement in fit quality. The SNe+BAO combination stands out as the only pairwise case in which both criteria concur in favoring the parametric model ($\Delta\mathrm{AIC} = -8.537$, $\Delta\mathrm{BIC} = -3.093$), with $\Delta\mathrm{AIC}$ reaching the ``substantial'' range on the Akaike-weight scale \cite{burnham2007model},. This result identifies the SNe+BAO combination as the principal source of the preference for the parametric $H(z)$ model that persists in the full CC+BAO+SNe fit: incorporating CC into this combination leaves the AIC preference essentially unchanged ($-8.537 \rightarrow -7.006$) while further diluting the already weak BIC preference ($-3.093 \rightarrow -1.543$), consistent with CC contributing comparatively few data points relative to the additional penalization it imposes under the $k\ln N$ term.

We therefore interpret the results of Tables~\ref{tb5} and~\ref{tb6} as demonstrating that the preference for the parametric $H(z)$ model observed in the full combined analysis is not driven by any single dataset in isolation, but emerges specifically from the joint constraining power of SNe and BAO, with CC contributing comparatively little additional discriminating information in either direction. This pattern, a preference for the parametric model that appears only in combination and not in any individual dataset, is itself informative: it is consistent with a small, systematic deviation in $H(z)$ that becomes statistically accessible only once the degeneracy between parameters is broken by combining a geometric distance probe (BAO) with a luminosity-distance probe (SNe), rather than with a large or catastrophic failure of $\Lambda$CDM, which would be expected to manifest already at the level of the individual datasets. The fact that the signal is specifically tied to the SNe+BAO pairing, rather than simply increasing with the number of data points (CC+SNe does not show a comparable preference despite a similar combined sample size), further suggests that it is the complementary nature of the geometric and luminosity information, rather than statistical volume alone, that is responsible for the result.

Given the weak $\Delta\mathrm{BIC}$ obtained even for the full combination, we regard this result as a mild but consistent indication of residual structure in $H(z)$ rather than a definitive detection, warranting further investigation with higher-precision CC measurements and next-generation BAO and SNe compilations.

\section{Stability analysis by scalar perturbations}
\label{sec7}

In this section, we perform a stability analysis of the cosmological $f(T)$ framework by employing a scalar perturbation method. This investigation is crucial because the cosmological behavior relies on multiple simplifying assumptions, which obscures the general validity of the model. To confirm these findings, exploring the qualitative dynamics of the field equations becomes indispensable. Specifically, we consider linear, homogeneous, and isotropic perturbations to test the robustness of the derived cosmological solutions. Building on existing literature \cite{Wu:2012hs,Golovnev:2018wbh}, we model fluctuations in the Hubble parameter and energy density through the geometric and matter perturbation functions, $\delta(t)$ and $\delta_m(t)$, respectively. The perturbations are given by the following equations:
\begin{equation}
\label{h_pert}
    H(t) \rightarrow H(t)(1 + \delta),
\end{equation}
\begin{equation}
\label{ro_pert}
    \rho \rightarrow \rho (1 + \delta_m).
\end{equation}

Starting from the modified Friedmann equation, and noting that the modified action depends on the torsion scalar $T$, which in turn is related to the Hubble parameter $H$ by $T = -6H^2$, the perturbed torsion scalar expands to first order as:
\begin{equation*}
    T \rightarrow -6H^2(1 + \delta)^2 \approx -6H^2(1 + 2\delta) = T + 2T\delta.
\end{equation*}

Expanding $f(T)$ and $f_T(T) \equiv \frac{df}{dT}$ in a Taylor series around the background value of $T$, the first-order expressions read:
\begin{equation*}
\begin{aligned}
    f(T + 2T\delta) &\approx f(T) + 2T f_T \delta, \\
    f_T(T + 2T\delta) &\approx f_T(T) + 2T f_{TT} \delta.
\end{aligned}
\end{equation*}

Substituting these expansions into the modified Friedmann equation yields:
\begin{equation}
    \underbrace{3H^2 + T f_T - \frac{f}{2}}_{\text{Zero-order term}} + \left(6H^2 + T f_T + 2T^2 f_{TT}\right) \delta = \kappa^2 \rho (1 + \delta_m).
\end{equation}

By subtracting the zero-order background equation, we isolate the first-order perturbation relation:
\begin{equation}
    T (f_T + 2T f_{TT} - 1)\delta(t) = \kappa^2 \rho(t) \delta_m(t),
\end{equation}
which provides a direct connection between geometric and matter perturbations:
\begin{equation}
\label{delta_mod}
    \delta(t) = \frac{\kappa^2 \rho(t)}{T(t) \left[ f_T(T) + 2T f_{TT}(T) - 1 \right]} \delta_m(t).
\end{equation}

To close the system of perturbation equations, we utilize the energy-momentum conservation equation for cold dark matter (dust, $p=0$), given by $\dot{\rho} + 3H\rho = 0$. Perturbing this relation using equations \eqref{h_pert} and \eqref{ro_pert} leads to:
\begin{equation}
    \dot{\rho}(1 + \delta_m) + \rho \dot{\delta}_m + 3H\rho(1 + \delta + \delta_m) = 0.
\end{equation}

Applying the background conservation equation $\dot{\rho} = -3H\rho$, we obtain the evolution equation for the matter perturbation:
\begin{equation}
\label{delta_m_mod}
    \dot{\delta}_m(t) = -3H(t)\delta(t).
\end{equation}

Substituting Eq.~\eqref{delta_mod} into Eq.~\eqref{delta_m_mod} yields the differential equation for $\delta_m(t)$:
\begin{equation}
    \dot{\delta}_m(t) = \left( \frac{\kappa^2 \rho(t)}{2H(t) \left[ f_T(T) + 2T f_{TT}(T) - 1 \right]} \right) \delta_m(t),
\end{equation}
whose formal solution is given by:
\begin{equation}
\label{sol_pert}
    \delta_m(t) = \delta_m(t_0) \exp \left( \int_{t_0}^t \frac{\kappa^2 \rho(t')}{2H(t') \left[ f_T(T) + 2T f_{TT}(T) - 1 \right]} dt' \right).
\end{equation}
To evaluate the integrand in Eq.~\eqref{sol_pert}, we take the time derivative of the background Friedmann equation and substitute the continuity relation $\kappa^2 \dot{\rho} = -3H \kappa^2 \rho$, deriving the model identity:
\begin{equation}
\label{identity_rho}
    \kappa^2 \rho(t) = 2\dot{H}(t) \left[ f_T(T) + 2T f_{TT}(T) - 1 \right].
\end{equation}

Substituting Eq.~\eqref{identity_rho} into the integrand causes the model-dependent functional terms to vanish:
\begin{equation}
    \frac{\kappa^2 \rho(t')}{2H(t') \left[ f_T(T) + 2T f_{TT}(T) - 1 \right]} = \frac{\dot{H}(t')}{H(t')}.
\end{equation}

Consequently, the differential equation simplifies to $\frac{\dot{\delta}_m}{\delta_m} = \frac{\dot{H}}{H}$, yielding the exact analytical solution for matter perturbations:
\begin{equation}
\label{sol_deltam_exact}
    \delta_m(t) = \delta_m(t_0) \frac{H(t)}{H(t_0)}.
\end{equation}

Furthermore, substituting Eq.~\eqref{identity_rho} and Eq.~\eqref{sol_deltam_exact} back into Eq.~\eqref{delta_mod}, the geometric perturbation $\delta(t)$ simplifies to:
\begin{equation}
\label{sol_delta_exact}
    \delta(t) = -\frac{\dot{H}(t)}{3 H(t) H(t_0)} \delta_m(t_0).
\end{equation}

Equations~\eqref{sol_deltam_exact} and \eqref{sol_deltam_exact} demonstrate that the evolution of perturbations is dictated by the background expansion history. Since $\dot{H}(t) < 0$ throughout the cosmic expansion, the Hubble parameter $H(t)$ decays monotonically over time. As a result, both $\delta_m(t)$ and $\delta(t)$ decay proportionally to $H(t)$, confirming that the cosmological background solution is dynamically stable against linear, isotropic, and homogeneous fluctuations.

The qualitative behavior of these analytical solutions is depicted in Figs.~\ref{fig_delta} and \ref{fig_deltam}, evaluated using the constrained values for $H_0$, $\alpha$, and $\beta$ reported in Table~\ref{tb2}. Figure~\ref{fig_delta} illustrates the evolution of the Hubble perturbation $\delta(t)$ as a function of redshift $z$ and initial condition ratios $\delta_m(t_0)/H_0$, showing a smooth, monotonic decay toward the present era ($z=0$). Similarly, Fig.~\ref{fig_deltam} displays the 3D surface profile of the matter perturbation $\delta_m(t)$, confirming that the perturbation remains bounded and decays monotonically without developing wild oscillations or exponential instabilities. Provided that the regularity condition $f_T(T) + 2T f_{TT}(T) - 1 \neq 0$ is preserved across the entire cosmic history, these findings confirm the overall physical viability and dynamical robustness of the proposed $f(T)$ framework\cite{Mussatayeva:2023aoa, Duchaniya:2022rqu}.

\begin{center}
    \vspace{0.3cm}
    \includegraphics[width=0.6\linewidth]{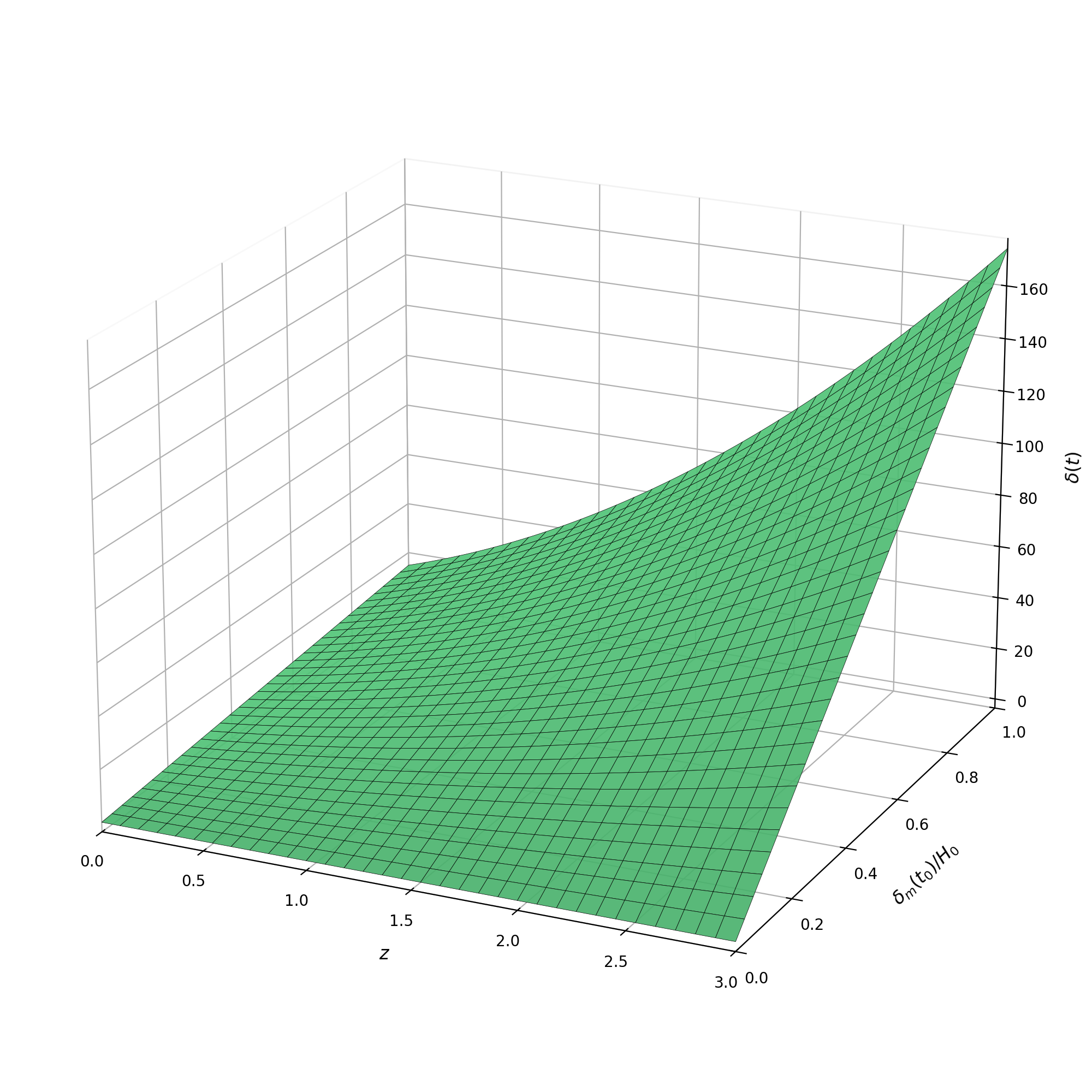}
    \captionof{figure}{Evolution of the Hubble perturbation parameter $\delta(t)$ as a function of redshift $z$ and the initial condition ratio $\delta_m(t_0)/H_0$. The surface demonstrates the monotonic decay of perturbations toward the present ($z=0$), corroborating the background dynamical stability of the $f(T)$ model.}
    \label{fig_delta}
    \vspace{0.3cm}
\end{center}

\begin{center}
    \vspace{0.3cm}
    \includegraphics[width=0.6\linewidth]{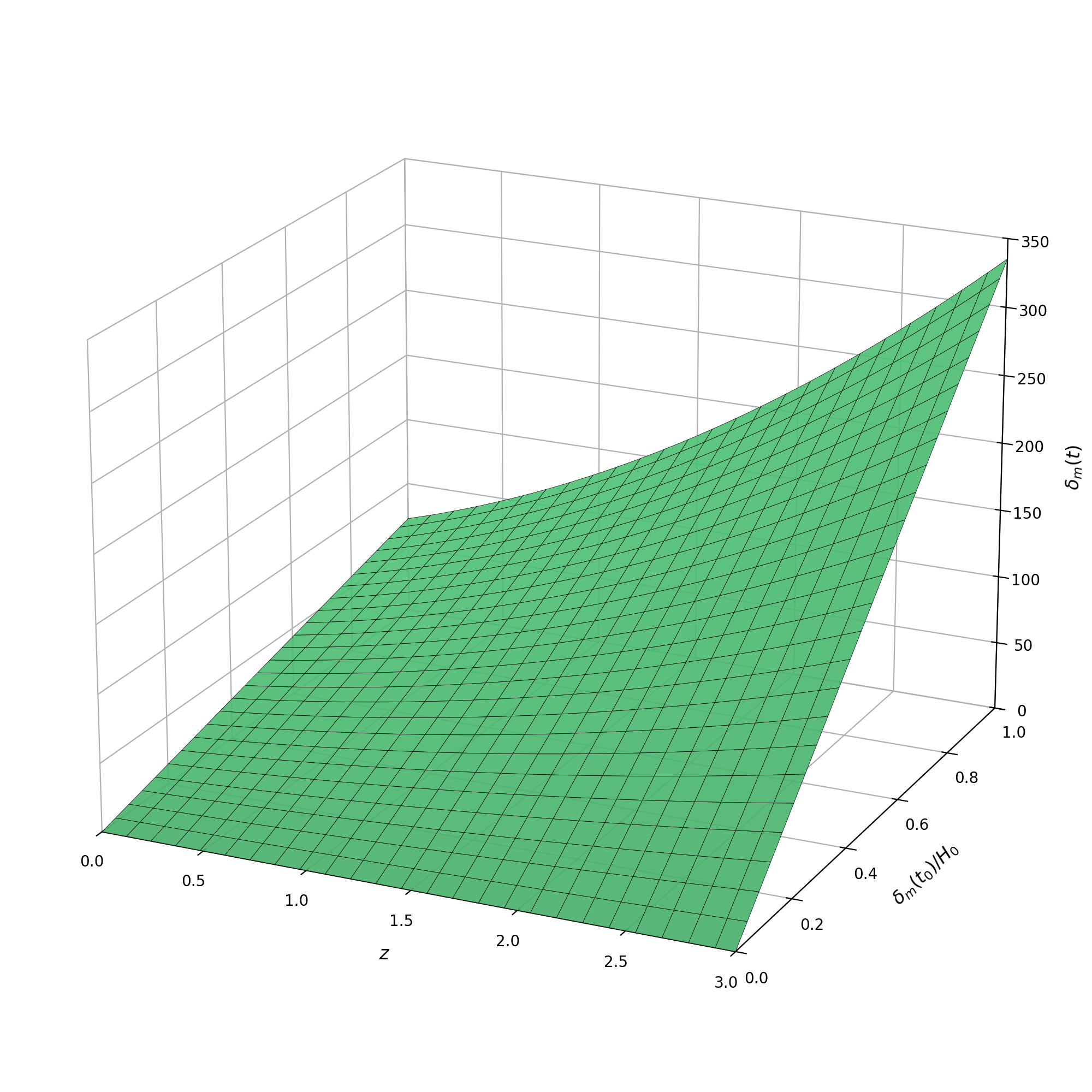}
    \captionof{figure}{Evolution of the matter perturbation parameter $\delta_m(t)$ as a function of redshift $z$ and the initial condition ratio $\delta_m(t_0)/H_0$. The smooth decay toward $z=0$ explicitly reflects the analytical relation $\delta_m(t) = \delta_m(t_0) H(t)/H_0$.}
    \label{fig_deltam}
    \vspace{0.3cm}
\end{center}

\section{Final remarks}

In this work we explore the possibility of a directly modified gravity model in the action of the type $T + f(T)$, in which the function $f(T)$ is constructed in such a way (see Eq.\eqref{lag1}) that in a specific case ($n = 0$), we return to the standard cosmological model with a cosmological constant $\Lambda$. To aid and validate the analysis of the model, we use a parameterization of $H(z)$ to test the free parameters $H_0$, $\alpha$ and $\beta$ against observational data of Supernovae Ia with the Pantheon+SH0ES data, DESI RD2 and of Cosmic Chronometers data. The test performed with the data was the maximization of likelihood and the best-fits found were $H_0 = 72.518^{+0.224}_{-0.230}$, $\alpha = 0.612^{+0.038}_{-0.038}$, $\beta = 0.198^{+0.015}_{-0.014}$ and $r_d = 136.125^{+1.263}_{-1.234}$, thus showing that the universe is going through a phase of accelerated expansion based on the behavior of the parameter $q$. The behavior of the model for the main quantities $\rho_{g}$, $p_{g}$ and $\omega_{g}$ were tested for different values of $n$, taking into account the best-fits mentioned in Table \ref{tb2}. We found that for the interval $-1\leq n\leq 1$, we can find solutions similar to the quintessence models for positive $n$ and phantom models for negative $n$, without loss of generality when we take the limit of $\omega_{g}$ for $z \rightarrow \infty$. We can find solutions similar to the quintessence models for positive $n$ and phantom models for negative $n$, without loss of generality when we take the limit of $\omega_{g}$ for $z \rightarrow \infty$. It is important to emphasize that the $H(z)$ parametrization employed here corresponds to a model-independent analysis technique, in the sense that it does not derive from a specific first-principles Lagrangian or a closed set of field equations, but rather from a flexible functional ansatz calibrated directly against the expansion-rate data. As such, the resulting best-fit values should be understood as a qualitative diagnostic of whether the data favor deviations from $\Lambda$CDM-like expansion history, and not as a direct solution or a first-principles prediction of the underlying $f(T)$ gravity model. The connection between the phenomenological $H(z)$ template and the modified gravity framework explored in this work is therefore indirect, serving primarily to validate the qualitative trends, such as the accelerated expansion and the quintessence/phantom-like behavior of $\omega_g$, rather than to establish a quantitative one-to-one correspondence between the fitted parameters $\alpha$, $\beta$ and the theoretical parameters of the $f(T)$ Lagrangian.

Regarding the statistical robustness of our results, the model selection analysis based on AIC and BIC provides a nuanced but consistent picture. The AIC comparison ($\Delta\text{AIC} = -7.006$) places our three-parameter $H(z)$ template close to the boundary of strong evidence against $\Lambda$CDM, suggesting that the improvement in fit quality is not merely a statistical artifact of the additional degree of freedom. The BIC, which penalizes model complexity more severely for large sample sizes, yields a more conservative assessment ($\Delta\text{BIC} = -1.543$), classifying the preference as weak on the Kass--Raftery/Jeffreys scale. This divergence between the two criteria is expected and does not indicate an inconsistency in our analysis; rather, it reflects the well-known trade-off between goodness-of-fit and parsimony that becomes more pronounced as $N$ grows. Taken together, we interpret these results as indicating that the extra parameter introduced by our phenomenological $H(z)$ parametrization is statistically justified by the data at a moderate level, without constituting decisive evidence against the standard cosmological model. This motivates further investigation with future, higher-precision datasets, which may help discriminate more sharply between the two scenarios.

Furthermore, we conducted a dynamical stability analysis of the cosmological background under linear, homogeneous, and isotropic scalar perturbations. By solving the perturbed field equations, we derived an exact analytical solution showing that both the matter density perturbation $\delta_m(t)$ and the Hubble parameter perturbation $\delta(t)$ decay monotonically over time, scaling directly with $H(t)$. This demonstrates that the cosmological background is dynamically stable against linear fluctuations, provided that the regularity condition $f_T(T) + 2T f_{TT}(T) - 1 \neq 0$ is preserved throughout the cosmic evolution.

Finally, we can state that both the parameterization of $H(z)$ and the model $f(T)$ are consistent with the observations, since the behavior presented in the graphs is within the expected and what has been found in the literature, despite not presenting a solution with positive values for pressure. However, here we are dealing with something effective and not with a fluid itself. In view of everything that was discussed in this work, we can reaffirm that modified gravity models with torsion can be a good alternative for studying new cosmological models that bring the necessary corrections to explain the open problem of accelerated cosmic expansion, whether momentary or not.

As a natural extension of this work, we plan to pursue in a future study a direct treatment of the logarithmic $f(T)$ model introduced here, obtaining an explicit solution of the modified Friedmann equations (Eqs.~(11)--(12)) rather than relying on the model-independent $H(z)$ parametrization employed in this analysis. Such a direct approach would allow $H_0$, $\Omega_{m,0}$ (or the analogous geometric density parameters), and the $f(T)$ parameters $\Lambda$, $n$, and $\lambda$ to be constrained jointly and self-consistently from the same observational datasets, rather than treating the kinematic fit and the gravitational sector as decoupled, as was necessarily done in the present cosmographic diagnostic.

\section*{Acknowledgments}
The authors thank the Conselho Nacional de Desenvolvimento Científico e Tecnológico (CNPq), Brazil (Grant No. 141416/2023-8) (AOA), and JEGSilva (Grant No. 310560/2025-0).

\section*{Data availability}
The observational datasets analyzed in this work are publicly available from their respective repositories:

\begin{itemize}
    \item \textbf{Cosmic Chronometers (CC):} The $H(z)$ measurements compiled by Moresco et al. are available in the original publications \cite{michele:01} and via public repositories (\url{https://gitlab.com/mmoresco/CCcovariance}).
    \item \textbf{Pantheon + SH0ES:} Distance moduli and covariance matrices \cite{Scolnic_2022} are available at \url{https://github.com/PantheonPlusSH0ES/DataRelease}.
    \item \textbf{DESI (DR2):} The BAO measurements are accessible via the Table IV from \cite{DESI:2025zgx}.
\end{itemize}

\bibliographystyle{unsrt}
\bibliography{refs}

\end{document}